\documentclass[a4paper,11pt]{article}
\usepackage{jheppub} 
\usepackage{lineno}
\usepackage{hyperref}
\usepackage{latexsym,bm,amsmath,amssymb,mathrsfs}
\usepackage{bbm}
\usepackage{physics}
\usepackage{color}
\usepackage{tikz}
\usepackage{tikz-feynman}
\usetikzlibrary{arrows}
\usepackage{graphicx}
\usepackage{multirow}
\usepackage{cancel}
\usepackage{diagbox}
\usepackage{makecell}

\newcommand{\br}{{\bm r}}
\newcommand{\bk}{{\bm k}}
\newcommand{\ba}{{\bm a}}
\newcommand{\bA}{{\bm A}}
\newcommand{\bq}{{\bm q}}
\newcommand{\bi}{{\mathrm{i}}}
\newcommand{\me}{{\mathrm{e}}}
\newcommand{\calN}{{\mathcal{N}}}

\newcommand{\sgn}{{\mathrm{sgn}}}
\newcommand{\nn}{\nonumber \\}

\title{\boldmath  
	Linear-T Resistivity from  Spatially Random Vector Coupling}

\author[a]{Yi-Li Wang,}
\author[b]{~Xian-Hui Ge,}
\author[a]{~Sang-Jin Sin}
\affiliation[a]{Department of Physics, Hanyang University,\\
	222 Wangsimni-ro, Seoul, 04763, Korea}
\affiliation[b]{ Department of Physics, Shanghai University,\\
	99 Shangda Rd., Shanghai, 200444, P.R. China}
\emailAdd{sjsin@hanyang.ac.kr}

\abstract{Recently, Patel \textit{et al.} introduced a higher dimensional version of the SYK model with 
	random coupling in a Yukawa interaction to find  the linear-$T$ resistivity. 
We test the universality of the mechanism by  replacing the scalar field with a vector field in various dimensions.  We find that   it works for vector and  scalar interactions, although the details are largely different. However,  this mechanism for the linear-$T$ resistivity works only in $(2+1)$ dimensions and not in  higher dimensions, regardless of the interaction type. 
Based on these results, we explore the r\^ole of spatial random disorder  and find a simple explanation of how such random scattering converts  the Fermi liquid to a strange metal by changing the self-energies  of the  involved  bosons and fermions. }
 
\begin{document}
\maketitle
\flushbottom

\section{Introduction}
\label{sec:intro}

 Understanding the mechanism of the strange metal  is considered as one of the most important topics in modern physics \cite{RevModPhys.78.17,Anderson_2017,Lee:2017njh,Greene2020,RevModPhys.92.031001,Hartnoll:2021ydi,Phillips:2022nxs}, because the strange metal appears ubiquitously in  the metallic phase of strongly interacting systems as well as the normal phase of the high-temperature superconductors.
 While the properties of strange metals are simple and universally characterized by the linear-$T$ resistivity,  there has not been any theoretical model to give such behavior.\\

The strange metal violates various features of the Landau's Fermi liquid, as observed in several systems, such as cuprate superconductors and heavy-fermion materials \cite{RevModPhys.92.031001,Phillips:2022nxs,PhysRevLett.59.1337,PhysRevLett.69.2975}. 
It is often accompanied by a logarithmic specific heat proportional to $T\ln(1/T)$ \cite{Hartnoll:2021ydi}, indicating the existence of the quantum criticality \cite{Lohneysen:2007zz,PMID:18097398,Shibauchi,Michon_2019}. Moreover, the resistivity coefficient $A$ in $\rho=\rho_0+A T$ is found to be closely related to the critical temperature $T_c$ such that $T_c\sim A^{\Box}$ \cite{Phillips:2022nxs}, where $A$ is also known to be correlated with superfluid density \cite{RevModPhys.78.17,Phillips:2022nxs}, and $\Box$ is a number to be fixed. All these observations were waiting for a theory of the strange metal, which is expected to be the key to comprehending the high-temperature superconductivity.  Despite the tremendous efforts over the past few decades, the strange metal remains mysterious. 
\\

Very  recently,  however, inspired by the Sachedev-Ye-Kitaev (SYK) model \cite{PhysRevLett.70.3339,Kitaev,Chowdhury2022}, the authors of \cite{Patel:2022gdh} attacked the problem using a spatially random coupling. They achieved the strange-metal behavior by randomizing a Yukawa interaction, $g\psi^{\dagger}\psi\phi$, between a Fermi surface (FS)  and  critically fluctuating scalar bosons  \cite{Lee:2017njh, Guo2022, Esterlis2021, Patel:2022gdh}, where $\psi$ and $\phi$ represent the electron and the scalar field respectively. The interaction reads $g_{ijk}(\br)\psi^{\dagger}_i\psi_j\phi_k$ with a space-dependent coupling constant, and $g_{ijk}(\br)$ has zero mean and non-vanishing correlation  $\langle g_{ijk}^*(\br)g_{i'j'k'}(\br')\rangle=g^2\delta(\br-\br')\delta_{ii'jj'kk'}$.  
This model bridged the chasm between the observations and the theoretical interpretation of strange metals for the first time.\\

It is of great interest to see how universal this mechanism is as a theory of strange metal. 
For this purpose, we extend the random Yukawa model in \cite{Patel:2022gdh} to a vector  coupling analogous to the QED, and calculate  its optical conductivity. In other words, we will consider an FS coupled to a vector field, where the coupling constant is spatially randomized. It will be shown  that the linear-$T$ resistivity still remains in this model, although the details are different.  \\

The rest of this article is organized as follows. In section \ref{sec:2d}, we build a system with a spatially random coupling between an FS and a vector field. After the corresponding `$G$-$\Sigma$' action is constructed following \cite{Esterlis2021,Guo2022,Patel:2022gdh,PhysRevB.63.134406,Sachdev:2015efa,Maldacena:2016hyu,Kitaev:2017awl,Sachdev2023}, we will show that such a system exhibits strange-metal behavior in $(2+1)$ dimensions. 
In section \ref{sec:3d},   we generalize our vector model as well as the scalar model \cite{Patel:2022gdh} to $(3+1)$-dimensional space-time, and show that  linear-$T$ resistivity is no longer there.  In section \ref{sec:discussion},  we explore the role of spatial random disorder and find a simple mechanism by which such a random scattering converts  the Fermi liquid to the strange metal by changing the propagators of bosons and fermions.  Conclusion and outlooks are presented in section \ref{sec:conclusion}. 

\section{The Critical Fermi Surface Randomly Coupled to a Vector Field}
\label{sec:2d}
\subsection{The Model and Saddle Point Equation}
Let us start with a two-dimensional spinless electron gas 
 coupled to a vector field. 
In Euclidean time $\tau$, the action of such a preliminary  model   reads \cite{Altland2023,PhysRevB.47.7312,Kim:1994tfo,Lee:2009epi,Chowdhury2022} :
	\begin{eqnarray}\label{eqn:2daction}
		S_{pre}&=& \int d\tau d^2\br \left[\psi^{\dagger}(\br,\tau)\left(\partial_\tau-\bi A^0-\frac{\nabla^2}{2m}-\mu\right)\psi(\br)
		+v(\br)\psi^{\dagger}(\br,\tau)\psi(\br,\tau)+\frac{1}{2}((\mathrm{\nabla}\times\bA )^2-\dot{A}^2)\right.\nonumber\\
		&&\left. +\frac{\bi e}{m}\psi^{\dagger}(\br,\tau)\nabla_a\psi(\br,\tau)\bA^a
		+\frac{e^2}{2m}\bA\cdot\bA\psi^{\dagger}(\br,\tau) \psi(\br,\tau)
		\right],
	\end{eqnarray}
	where $A^0$ is the scalar potential and $\bA=(A_x,A_y)$ is the vector potential. A spatial disorder is introduced by $v(\br)$, known as a potential disorder that features scattering impurities \cite{Patel:2022gdh,Guo2022,Altland2023,Coleman2019}. Here $m$ and $e$ respectively denote the bare mass and bare charge of an electron. Containing no topological term, this is not a suitable theory for studying the half-filled Landau level. However, as the simplest model one can have, (\ref{eqn:2daction}) offers a reasonable starting point for future generalization to more complicated and physically interesting systems. In action (\ref{eqn:2daction}), the time derivative is included to make the action as general as possible, though this part will become subdominant during the   computation \cite{Chowdhury2022}.\\

We follow the strategy suggested in \cite{Esterlis2021,Guo2022,Patel:2022gdh}, and introduce spatial randomness to the interaction between the FS and the vector field $A$. In order to trace the effects of random coupling, we replace $e$ with $K$ and 
 $e^2$ with $\tilde{K}$. We treat $K$ and $\tilde{K}$ as two independent parameters and randomize them separately. We take the result as our model and consider an \emph{external} electromagnetic field $\mathscr{A}$. 
Then, inspired by the action (\ref{eqn:2daction}), we can write an action 
	\begin{eqnarray}\label{eqn:action1}
		S&=& \int d\tau d^2\br \left[\sum_{ijl} \psi^{\dagger}_i(\br,\tau)\left(\partial_\tau+\frac{(-\bi\nabla+\mathscr{A})^2}{2m}-\mu\right)\psi_i(\br)
		+\frac{1}{\sqrt{N}}\sum_{ij} v_{ij}(\br)\psi^{\dagger}_i(\br,\tau)\psi_j(\br,\tau)\right.\nonumber\\
		&& -\frac{\bi}{N}\sum_{ijl}K_{ijl}(\br)\psi^{\dagger}_i(\br,\tau) \psi_j(\br,\tau) A_l
		+\frac{1}{N}\frac{\bi}{m}\sum_{ijl}K_{ijl}(\br)\psi^{\dagger}_i(\br,\tau)\nabla_{a}\psi_j(\br,\tau)\bA^{a}_l\nn
		&& -\frac{1}{N}\frac{1}{m}\sum_{ijl}\check{K}_{ijl}(\br)\psi^{\dagger}_i(\br,\tau)\psi_j(\br,\tau)\bA^{a}_l\mathscr{A}_a
		+\frac{1}{N^{3/2}}\sum_{ijst}\frac{\tilde{K}_{ijst}(\br)}{2m}\bA_s\cdot\bA_t\psi_i^{\dagger} \psi_j\nn
		&&\left.+\frac{1}{2}(A^{a}(g_{ab}\Box)A^{b})
		\right].
\end{eqnarray}
The gauge-invariant action \eqref{eqn:2daction} is the ancestor action that we start from, but the gauge invariance is now broken due to the randomization, where $K$ and $\tilde{K}$ are treated as two independent parameters. Moreover, the kinetic kernel of $U(1)$ gauge fields has no inverse in general \cite{Ryder_1996}, so gauge-fixing is necessary to get the propagator. It is convenient to choose a \emph{Coulomb gauge} such that $\mathrm{\nabla}\cdot\bA=0$ \cite{Chowdhury2022,Lee:2009epi,Kim:1994tfo,Ryder_1996,Mahan2014,Coleman2019}, so the spatial part of the gauge field is transverse. Furthermore, for simplicity, the scalar potential 
set to be $A^0=0$ assuming that there is no external charge source. In our case, however, without gauge invariance, no gauge-fixing can be performed. Here we directly choose a kinetic kernel $(-\partial_\tau^2-\nabla^2)$ for the vector field and require also $A^0=0$ to avoid the complication. As a result, this field has two spatial degrees of freedom, and each component is a scalar boson. Hence, its dynamics is similar to a $U(1)$ gauge field with Coulomb gauge and temporal gauge, but we should emphasize again that it is a vector field instead of a gauge field since there is no gauge invariance anymore.\\

Following  \cite{Esterlis2021,Guo2022,Patel:2022gdh}, we assume that 
  $v_{ij}$, $K_{ij}$, and $\tilde{K}_{ij}$ obey the Gau\ss ian distribution with zero mean  and  satisfy 
\begin{eqnarray}
	&&\langle v_{ij}^*(\br)v_{i'j'}(\br')\rangle=v^2 \delta(\br-\br')\delta_{ii'}\delta_{jj'},\\
	&&\langle K_{ijl}^*(\br)K_{i'j'l'}(\br')\rangle=K^2 \delta(\br-\br')\delta_{ii'}\delta_{jj'}\delta_{ll'},\\
	&&\langle \tilde{K}_{ijst}^*(\br)\tilde{K}_{i'j's't'}(\br')\rangle=\tilde{K}^2 \delta(\br-\br')\delta_{ii'}\delta_{jj'}\delta_{ss'}\delta_{tt'}.
\end{eqnarray}
The field $A^{\mu}$ is also labeled by a flavor ($i=1,...,N$) so that fluctuations around the saddle point is well controlled at large-$N$ limit \cite{Esterlis2021,Guo2022,Patel:2022gdh}. Here we label $\bA^{a}\psi^{\dagger}\nabla_{a}\psi$ and $A_{\mu}\mathscr{A}^{\mu}\psi^{\dagger}\psi$ by two independent parameters $K_{ijl}$ and $\check{K}_{ijl}$, though they share the same norm. This means that we randomize each single vertex separately. An alternative randomization is using $K_{ijl}$ also for  $A_{\mu}\mathscr{A}^{\mu}\psi^{\dagger}\psi$. As we will show later, both two choices are able to produce linear resistivity, but the current choice will make the computation much simpler.\\

Next,  we  need a  `$G$-$\Sigma$' theory to study the large-$N$ limit of the theory \cite{Esterlis2021,Guo2022,Patel:2022gdh,PhysRevB.63.134406,Sachdev:2015efa,Maldacena:2016hyu,Kitaev:2017awl,Sachdev2023}. To write the action (\ref{eqn:action1}) in a standard form that describes an FS coupled to an emergent vector field, we rescale $K A_{\mu}= a_{\mu}$ \cite{PhysRevB.42.10348,PhysRevB.37.580,PhysRevLett.65.653,PhysRevLett.60.821,Wiegmann1988,PhysRevB.47.7312,Kim:1994tfo}.
Let $\Sigma/\Pi^{\mu\nu}$ and $G/D^{\mu\nu}$ denote the self-energy/polarization  and propagator of the fermion/boson field. Then optical conductivity can be directly read from the polarization tensor of an external gauge field through the Kubo formula. The key idea of $G$-$\Sigma$ theory is to introduce a set of bi-local variables 
$$G(x_1,x_2), \quad D^{\mu\nu}(x_1,x_2), \quad \Sigma(x_1,x_2) , \quad \Pi^{\mu\nu}(x_1,x_2), $$ 
	and these will yield the propagators and self-energies at the saddle point where 
	$$\delta S/\delta G=\delta S/\delta \Sigma=\delta S/\delta D^{\mu\nu}=\delta S/\delta \Pi^{\mu\nu}=0, $$
	 with $x=(\tau,\br)$. 
The bi-local variables $G$ and $D$ are defined as follows,
\begin{eqnarray}
	&&G(x_1,x_2)\equiv-\frac{1}{N}\sum_{i}\langle\mathcal{T}\left(\psi_i(x_1)\psi_i^{\dagger}(x_2)\right)\rangle,\\
	&&D^{\mu\nu}(x_1,x_2)\equiv\frac{1}{N}\sum_{i}\langle\mathcal{T}\left(a_i^{\mu}(x_1)a_i^{\nu}(x_2)\right)\rangle.
\end{eqnarray}
Together with these bi-local variables, $G(x_1,x_2)$ and $D^{\mu\nu}(x_1,x_2)$, the `self-energies' $\Sigma, \Pi_{\mu\nu}$ are added as Lagrangian multipliers. To derive the saddle point equation, let us temporally take the external field $\mathscr{A}\to 0$, which is of no consequence at this stage. The $G$-$\Sigma$ action thus reads
\begin{eqnarray}\label{eqn:action3}
	S&=& \int d\tau d^2\br \left[\sum_{ijl} \psi^{\dagger}_i(\br,\tau)\left(\partial_\tau-\frac{\nabla^2}{2m}-\mu\right)\psi_i(\br,\tau)
	+\frac{1}{\sqrt{N}}\sum_{ij} v_{ij}(\br)\psi^{\dagger}_i(\br,\tau)\psi_j(\br,\tau)\right.\nonumber\\
	&&\left.+\frac{1}{2K^2} a^{\mu}\left(-\partial_\tau^2+\bq^2\right)a^{\nu}
	+\sum_{ijl}\frac{1}{KN}K_{ijl}\frac{\bi}{m}\psi^{\dagger}_i\nabla_a\psi_j\ba_l^a
	+\frac{1}{K^2N^{3/2}}\sum_{ijst}\frac{\tilde{K}_{ijst}}{2m}\ba_s\cdot\ba_t\psi_i^{\dagger} \psi_j
	\right]\nonumber\\
	&&-N\int d\tau d\tau' d^3\br \Sigma \left(G+\frac{1}{N}\sum_i \psi_i\psi^{\dagger}_i\right)
	+\frac{N}{2}\int d\tau d\tau' d^3\br \Pi^{\mu\nu}\left(D^{\mu\nu}-\frac{1}{N}\sum_i a_i^{\mu}a_i^{\nu}\right). 
\end{eqnarray}
Performing disorder average via the replica trick \cite{Altland2023} using the fact   \cite{Altland2023} 
 \[ 
	\int d(\psi,\psi^{\dagger})e^{-(\psi^{\dagger})^T\mathrm{M}\psi}=\det(\mathrm{M}),\]
for generic complex matrices $M$,
 one   gets
\begin{eqnarray}\label{eqn:action2}
	\frac{S}{N}&=&-\ln \det((\partial_\tau+\varepsilon_k-\mu)\delta(x-x')+\Sigma)\nonumber\\
	&&+\frac{1}{2}\ln\det(-\frac{g_{ab}}{K^2}(-\partial_\tau^2+\bq^2)\delta(x-x')-\Pi_{ab})\nonumber\\
	&&+\Tr \left(\frac{v^2}{2}G\cdot G \bar{\delta}\right)
	+\frac{1}{2m^2}\Tr\left(\frac{(k_1+k_2)^a(k_1+k_2)^b}{4}G(k_1) D_{ab}\cdot G(k_2)\bar{\delta}\right)\nonumber\\
	&&+\Tr \left(\frac{1}{8m^2}G D_{ab}D^{ab} \bar{\delta} G\right)
	-\Tr(\Sigma\cdot G)+\frac{1}{2}\Tr(\Pi^{ab}D_{ab}),
\end{eqnarray}
where $\bk$ is the momentum of fermions and $\bar{\delta}$ is the spatial delta function coming from the disorder average. The labels $a,b=1,2$ represent the spatial components. Since $A^0=0$, $D_{00}$ and $\Pi_{00}$ are neglected. Here $\varepsilon_{\bk}$ is the fermion spectrum $\omega=\varepsilon_{\bk}=\bk^2/(2m)$ \cite{Coleman2019}. We use the following short-handed notation 
 \cite{Guo2022,Gu:2019jub},
\begin{equation}
	\Tr(f_1\cdot f_2)\equiv f_1^Tf_2\equiv\int dx_1dx_2f_1(x_2,x_1)f_2(x_1,x_2),
\end{equation}
where the transpose reads
$	f^T(x_1,x_2)\equiv f(x_2,x_1).$
Finally, starting from the action (\ref{eqn:action2}), one derives the saddle-point equation 
\begin{eqnarray}
	\frac{\delta S}{N}
	&=&\Tr \delta \Sigma\left( \left(-\partial_\tau-\varepsilon_k+\mu-\Sigma\right)^{-1}-G\right)\nonumber\\
	&&+\frac{1}{2}\Tr \delta D^{ab}\left(\Pi_{ab}+\frac{1}{m^2}\frac{(k_1+k_2)_a(k_1+k_2)_b}{4}G\cdot G\bar{\delta}
	+\frac{1}{4m^2}G D_{ab}G\bar{\delta}\right)\nonumber\\
	&&+\Tr \delta G \left(-\Sigma+
	v^2 G\bar{\delta}+\frac{1}{m^2}\frac{(k_1+k_2)^a(k_1+k_2)^b}{4}D_{ab} G\bar{\delta}+\frac{1}{4m^2}D_{\mu\nu}D^{\mu\nu}G\bar{\delta}\right) \nonumber\\
	&&+\frac{1}{2 }\Tr \delta \Pi^{ab}(D_{ab}-K^2(-(-\partial_\tau^2+\bq^2)g^{ab}-K^2\Pi^{ab})^{-1})\\
	&\equiv&
	\Tr(\delta\Sigma(G_*[\Sigma]-G)+\delta G(\Sigma_*[G]-\Sigma) 
+\frac{1}{2}\delta\Pi_{ab}(D^{ab}-{D_*}^{ab}[\Pi^{ab}])+\delta D_{ab}(\Pi^{ab}-\Pi_*^{ab}[D_{11}])),\nonumber
\end{eqnarray}
where $k_{\mu}=(0,\bk)$. 
One then obtains the Dyson equations
\begin{align}
	&G=G_*=\left(-\partial_\tau-\varepsilon_k+\mu-\Sigma\right)^{-1},\label{eqn:f-propagator}\\
	&\Sigma=\Sigma_*=
	v^2 G(\tau,\br=0)\delta^3(\br)+\frac{(k_1+k_2)^a(k_1+k_2)^b}{4m^2}D_{ab} G\bar{\delta}+\frac{1}{4m^2}D_{ab}D^{ab}G\bar{\delta},\label{eqn:f-selfenergy}\\
	&D_{ab}={D_*}_{ab}=K^2(-g^{ab}(-\partial_\tau^2+\bq^2)-K^2\Pi^{ab})^{-1},\label{eqn:b-propagator}\\
	&\Pi_{ab}={\Pi_*}_{ab}=-\frac{(k_1+k_2)_a(k_1+k_2)_b}{4m^2}G\bar{\delta}\cdot G
	-\frac{1}{4m^2}G D_{ab}G\bar{\delta},\label{eqn:b-selfenergy}
\end{align}
which take similar forms with those found in \cite{Kim:1994tfo}, and they can be directly evaluated via Feynman diagrams. We consider low-frequency fermion self-energies \cite{Esterlis2021,Guo2022,Patel:2022gdh} up to one-loop corrections \cite{Kim:1994tfo}. The Feynman diagram corresponding to the $DDG$ term in eqn. (\ref{eqn:f-selfenergy})
\begin{equation}\label{graph:ddg}
	\includegraphics[scale=0.4]{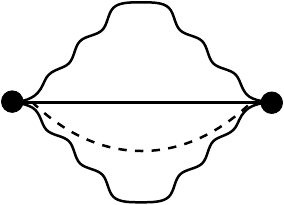}\,,
\end{equation}
where the dashed line represents the disorder average, can be neglected, since it is quadratic in frequency. Here, the solid line and the wavy line represent the electron propagator and the boson propagator respectively \footnote{Due to the spatial delta $\bar{\delta}$ in self-energies (\ref{eqn:f-selfenergy}) and (\ref{eqn:b-selfenergy}), the vertices of the propagators in these diagrams should be contracted. Here we use the ordinary Feynman diagram for clarity.}.\\

For the same reason, the terms $GGD$ in eqn. (\ref{eqn:b-selfenergy}), which corresponds to the following Feynman diagram 
\begin{equation}\label{graph:ggd}
	\includegraphics[scale=0.3]{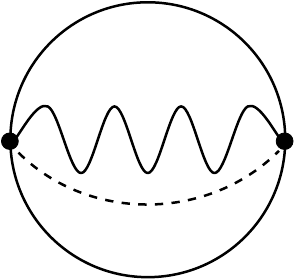}, 
\end{equation}
  is also negligible. However, when we compute the conductivity later after coupling the system with an external electromagnetic field $A_{\mu}$, this diagram is the key to linear resistivity, as will be shown later. 
In the presence of an external field, the polarization \eqref{graph:ggd} comes from $A_{\mu}a^{\mu}\psi^{\dagger}\psi$, which yields higher-order corrections in frequencies in \cite{Kim:1994tfo}, and this diagram does not exist in \cite{Patel:2022gdh}.  In our model, this polarization is not of higher order, so it will be computed as well.\\

Additionally, due to the fact that  $\langle\tilde{K}_{ij}\rangle=0$, there is neither one-loop corrections to the fermion self-energy  coming from 
\begin{equation}\label{eqn:self-energy-loop}
	\includegraphics[scale=0.25]{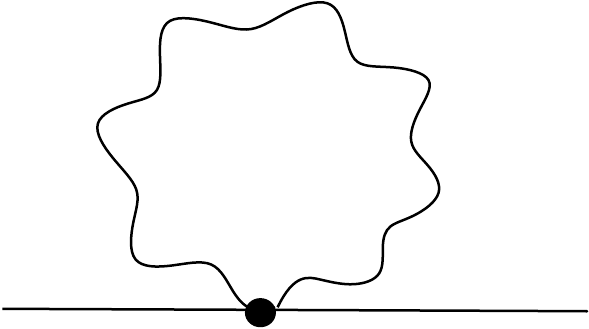}\,,
\end{equation}
nor the bosonic vector field polarisations  coming from 
\begin{equation}\label{eqn:polarisation-loop}
	\includegraphics[scale=0.2]{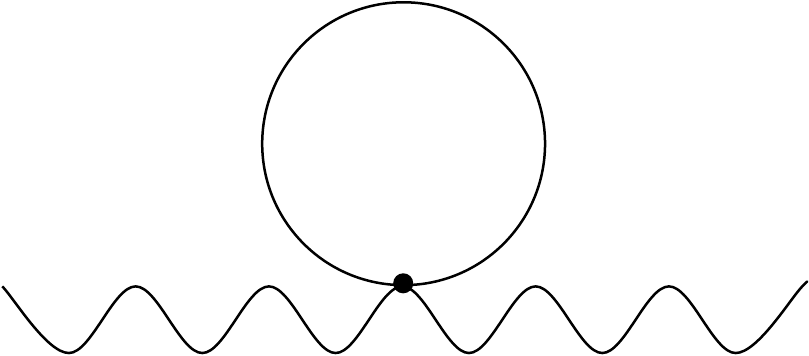}\,
\end{equation}
in our theory. This is a notable difference from the system considered in \cite{Kim:1994tfo}.

\subsection{Conductivity}

In order to find the conductivity, which is given by the polarization bubbles of the external electromagnetic field $\mathscr{A}_{\mu}$, we need to first find the propagators and the self-energies of electrons and bosons.\\

Due to the spatial delta $\bar{\delta}$ in eqn.(\ref{eqn:f-selfenergy}) and eqn.(\ref{eqn:b-selfenergy}), the momentum dependence of the self-energies is lost, i.e. $\Sigma(\bi\omega_n,\bk)=\Sigma(\bi\omega_n)$ and $\Pi_{ab}(\bi\Omega_m,\bq)=\Pi_{ab}(\bi\Omega_m)$, with $\omega_n, \Omega_m$ being the Matsubara frequencies.  We begin with the evaluation of the electron self-energy $\Sigma$, and define $\xi_{\bk}\equiv \varepsilon_{\bk}-\mu$. 
The dominant part of electron self-energy comes from the potential disorder (like that in \cite{Guo2022,Patel:2022gdh}), which reads
\begin{eqnarray}
	\Sigma_v(\bi\omega_n)&=&v^2\int\frac{d^3 \bk'}{(2\pi)^3}G(\bi\omega_n,\bk')\nonumber\\
	&=&v^2\int \mathcal{N} d\xi_{\bk'}\frac{1}{\bi\omega_n-\xi_{\bk'}-\Sigma(i\omega_n,\bk')}\nonumber\\
	&=&-\bi\frac{\Gamma}{2}\sgn(\omega_n),
\end{eqnarray}
where  $\Gamma\equiv2\pi v^2 \mathcal{N}$ is the disorder scattering rate, and $\mathcal{N}=m/(2\pi)$ is the Density of State (DoS) at the Fermi energy in two (spatial) dimensions. Here the integration over momentum $\int d\bk$ can be replaced by the energy integral $\int d\xi_{\bk}$ because the main contribution is from the electrons near the FS \cite{Coleman2019}.
It can be verified later that the typical peak of the electron propagator is wider than that of the bosonic vector field, so we do not simplify the calculation via a Prange-Kadanoff reduction used in \cite{Guo2022} on the saddle point. \\

Eqn.\eqref{eqn:f-selfenergy} implies that the computation of fermion self-energies requires first finding the boson propagator. To this end, we need to find the boson self-energy \eqref{eqn:b-selfenergy}.
Due to the spatial delta, the vertices $w$ and $w'$ in usual bosonic contracts, and one obtains a graph with one vertices and two loops as follows,
\begin{equation}
	\includegraphics[scale=0.5]{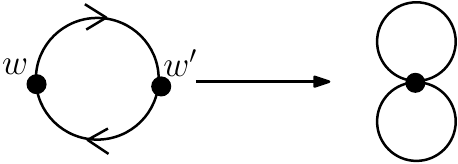}\,.
\end{equation}
As a result, the bosonic self-energy reads
\begin{eqnarray}\label{eqn:seboson}
	&&-K^2\Pi_0^{ab}(i\Omega_m)\nn
	&=&\delta^{ab}\frac{K^2}{4m^2}\calN^2\int\frac{d\omega_n}{2\pi}\int d\xi_{\bk}d\xi_{\bk'}
	\frac{k_F^2}{i\frac{\Gamma}{2}\sgn(\omega_n)-\xi_{\bk}}
	\left(\frac{1}{i\frac{\Gamma}{2}\sgn(\omega_n+\Omega_m)-\xi_{\bk'}}
	-\frac{1}{i\frac{\Gamma}{2}\sgn(\omega_n)-\xi_{\bk'}}\right)\nn
	&=&-\frac{i}{2}\delta^{ab}K^2 \frac{v_F^2}{4} \calN^2\delta^{\mu\nu}\int d\omega_n\int d\xi_{\bk'}\sgn(\omega_n)
	\left(\frac{1}{i\frac{\Gamma}{2}\sgn(\omega_n+\Omega_m)-\xi_{\bk'}}
	-\frac{1}{i\frac{\Gamma}{2}\sgn(\omega_n)-\xi_{\bk'}}\right)\nn
	&=&-\frac{\pi}{2}\delta^{ab}K^2\frac{v_F^2}{4}\calN^2	
	\int d\omega_n\sgn(\omega_n)\left(\sgn(\omega_n+\Omega_m)-\sgn(\omega_n)\right)\nn
	&=&\pi\calN^2K^2\frac{v_F^2}{4}|\Omega_m|\delta^{ab}\nn
	&\equiv& c_0|\Omega_m|\delta^{ab},
\end{eqnarray}
with $v_F$ the Fermi velocity. As we are interested in low temperatures, the Matsubara summation $T\sum_{\omega_n}$ is interchangeable with the integral $\int d\omega_n/(2\pi)$ \cite{Esterlis2021,Guo2022,Patel:2022gdh,Altland2023}.\\

To find the explicit form of boson propagators, assume 
\begin{equation}
	D_{ab}=K^2(-(-\partial_\tau^2+\bq^2)g^{ab}-K^2\Pi^{ab})^{-1}=-\frac{g_{ab}K^2}{(-\partial_\tau^2+\bq^2)\left(1+\Pi\right)}.
\end{equation}
By requiring 
\begin{equation}
	-\frac{g_{ab}}{(-\partial_\tau^2+\bq^2)\left(1+\Pi\right)}	(-(-\partial_\tau^2+\bq^2)g^{bc}-K^2\Pi^{bc}\bar{\delta})=\delta_a^c,
\end{equation}
one finds
\begin{equation}
	\Pi=\frac{c_0|\Omega_m|}{-\partial_\tau^2+\bq^2},
\end{equation}
so
\begin{equation}
	D_{ab}(i\Omega_m)=-\frac{g_{ab}K^2}{-\partial_\tau^2+\bq^2+c_0|\Omega_m|},
\end{equation}

Now let us turn to the electron self-energy. 
The one-loop correction to the electron self-energy from eqn. (\ref{eqn:f-selfenergy}) is graphically represented by
\begin{equation}
	\includegraphics[scale=0.4]{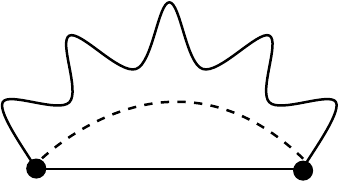}\,.
\end{equation}
So, there is also an contraction of vertex $w$ and $w'$ as illustrated below,
\begin{equation}
	\includegraphics[scale=0.5]{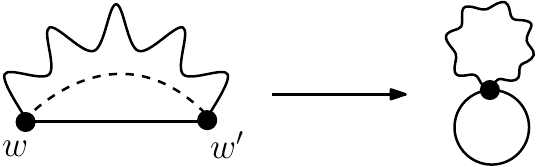}\,.
\end{equation}
This self-energy reads
\begin{eqnarray}\label{eqn:e-self-energy-oneloop}
	\Sigma_K(\bi\omega_n)&=&
	\frac{2K^2}{4m^2}\int\frac{d\Omega_m}{2\pi}\int\frac{d^2\bq}{(2\pi)^2}d\xi_{\bk}
	\frac{k_F^{2}}{\bi\frac{\Gamma}{2}\sgn(\omega_n+\Omega_m)-\xi_{\bk}}\frac{1}{\bq^2+c_0|\Omega_m|}\nn
	&=&-\bi v_F^2K^2\frac{\calN}{4}\int_{-\infty}^{+\infty}d\Omega_m\int_0^{\infty}\frac{|\bq|d|\bq|}{2\pi}
	\sgn(\omega_n+\Omega_m)\frac{1}{|\bq|^2+c_0|\Omega_m|}\nn
	&=&-\bi K^2\frac{v_F^2}{16\pi}\calN \int d\Omega_m\sgn(\omega_n+\Omega_m)\ln(\frac{\Lambda_q^2}{c_0|\Omega_m|})\nn
	&=& -\bi K^2 v_F^2\frac{\calN}{8\pi}\omega_n\ln(\frac{\me\Lambda_q^2}{c_0|\omega_n|}),
\end{eqnarray}
where $\Lambda_q$ is the UV cut-off on $\bq$ and $\me$ is the Euler's number (which should not be confused with the charge $e$). Therefore, we obtained a marginal-Fermi-liquid self-energy $\sim\omega\ln(1/\omega)$ \cite{Patel:2022gdh,PhysRevLett.69.2975}. This self-energy corresponds to a specific heat $\sim T\ln(1/T)$, which is one feature of the strange metal. Let us emphasis again that there is no momentum conservation imposed on the vertices, which can be seen from eqn.(\ref{eqn:e-self-energy-oneloop}). \\

We are now ready to add the external field $\mathscr{A}$ back and find the current-current correlator (the polarization).
The leading-order polarization comes from square of the $\mathscr{A}_{a}\psi^{\dagger}\nabla^{a}\psi$ coupling,
\begin{equation}\label{graph:jj}
	\includegraphics[scale=0.7]{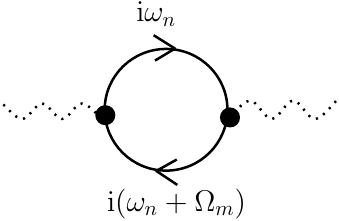}\,,
\end{equation}
where the dotted wavy lines represent the external gauge field line.
Due to the spatial delta, the polarizations and the boson self-energies are completely different. In order to distinguish these two quantities, let us denote the polarization as $\tilde{\Pi}_{\mu\nu}$. The basic polarization \eqref{graph:jj} is
\begin{eqnarray}\label{eqn:gauge_e_se}
	\tilde{\Pi}_{0}^{ab}(\bi\Omega_m)
	&=&\delta^{ab}\frac{1}{2m^2}\calN\int\frac{d\omega_n}{2\pi}\int d\xi_{\bk}
	\frac{k_F^2}{\bi\omega_n+\bi\frac{\Gamma}{2}\sgn(\omega_n)-\xi_{\bk}}\nn
	&&\times\left(\frac{1}{\bi(\omega_n+\Omega_m)+\bi\frac{\Gamma}{2}\sgn(\omega_n+\Omega_m)-\xi_{\bk}}
	-\frac{1}{\bi\omega_n+\bi\frac{\Gamma}{2}\sgn(\omega_n)-\xi_{\bk}}\right)\nn
	&=&\bi \delta^{ab} \frac{v_F^2}{2}\calN\frac{1}{2}\int d\omega_n
	\frac{\sgn (\omega_n)-\sgn(\omega_n+\Omega_m)}{\bi\frac{\Gamma}{2}\sgn(\omega_n)-\bi\frac{\Gamma}{2}\sgn(\omega_n+\Omega_m)-\bi\Omega_m}\nn
	&=&\delta^{ab}\frac{v_F^2}{2}\calN \frac{1}{\Gamma-\Omega_m} |\Omega_m| \equiv\frac{n_e }{m}\frac{1}{\Gamma-\Omega_m} |\Omega_m| \delta^{ab}, 
\end{eqnarray}
where 
$n_e$ is the equilibrium density of of electrons defined as  $\calN v_F^2/2=n_e/m$. 
Notice that there is no summation over spins because the fermions in this model are spinless. Eqn.(\ref{eqn:gauge_e_se}) describes the simplest contribution to the current-current correlator and yields a Drude-like form after one applies the Kubo formula \cite{Guo2022}.

At low frequencies, 
\begin{equation}\label{eqn:v-polar}
    \tilde{\Pi}_0^{11}(\bi\Omega_m)=\frac{n_e}{m}\frac{1}{\Gamma} |\Omega_m|+\mathcal{O}(\Omega_m^2),
\end{equation}
which is similar to the results in \cite{Patel:2022gdh}.\\

The Feynman diagrams representing the contribution of eqn.(\ref{eqn:e-self-energy-oneloop}) to the polarizations are  
\begin{equation}\label{graph:self1}
	\includegraphics[scale=0.45]{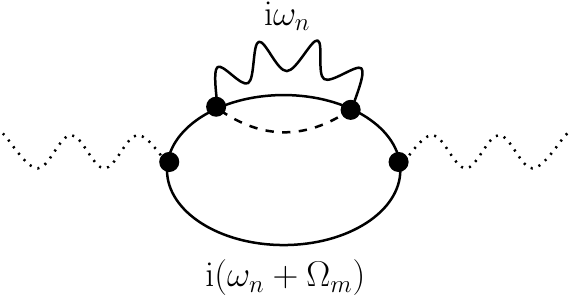}\,, \hskip 1 cm 
		\includegraphics[scale=0.45]{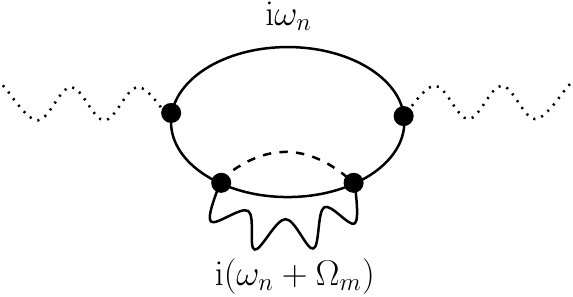}\,.
\end{equation}
The two diagrams bring the same contribution \cite{Patel:2022gdh}, which reads
\begin{eqnarray}\label{eqn:pi1}
	&&\tilde{\Pi}_{K(2)}^{ab}(\bi\Omega_m)=\tilde{\Pi}_{K(1)}^{ab}(\bi\Omega_m)\nn
	&=&
	\frac{1}{m^2}\calN\int\frac{d\omega_n}{2\pi}\int d\xi_{\bk} k^{a}k^{b}
	\left(\frac{1}{\bi\frac{\Gamma}{2}\sgn(\omega_n)-\xi_{\bk}}\right)^2\frac{1}{\bi\frac{\Gamma}{2}\sgn(\omega_n+\Omega_m)-\xi_{\bk}}\Sigma_K(\bi\omega_n)\Big|_{\Omega_m=0}^{\Omega_m}\nn
	&=&\bi\delta^{ab}\frac{K^2}{4}v_F^2\mathcal{N}\int d\omega_n
	\frac{-\sgn(\omega_n+\Omega_m)+\sgn(\omega_n)}{\left(\bi\frac{\Gamma}{2}\sgn(\omega_n+\Omega_m)-\bi\frac{\Gamma}{2}\sgn(\omega_n)\right)^2}
	\left[ -\bi v_F^2\omega_n\frac{\calN}{8\pi}\ln(\frac{\me\Lambda_q^2}{c_0|\omega_n|})\right]\nn
	&=&-\frac{K^2}{64\pi}v_F^4\frac{\calN^2}{\Gamma^2}
	\Omega_m^2\ln(\frac{\me^3\Lambda_q^4}{c_0^2|\Omega_m|^2})\delta^{ab}.
\end{eqnarray}
In addition, there are   vertex corrections coming from the Maki-Thompson diagrams (MT) diagrams, 
\begin{equation}\label{graph:mt}
	\includegraphics[scale=0.45]{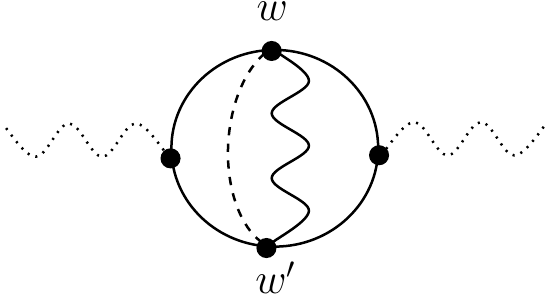}\,,
\end{equation}
and  Aslamazov-Larkin (AL) diagrams 
\begin{equation}\label{graph:al1}
	\includegraphics[scale=0.4]{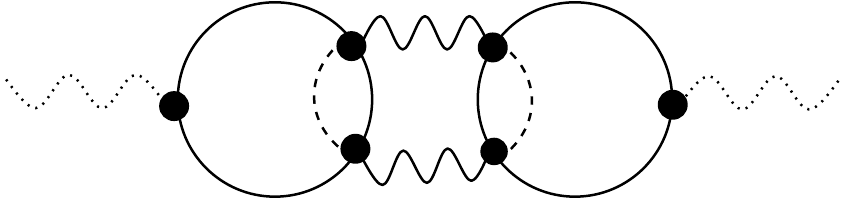}\, , \quad 
	\includegraphics[scale=0.4]{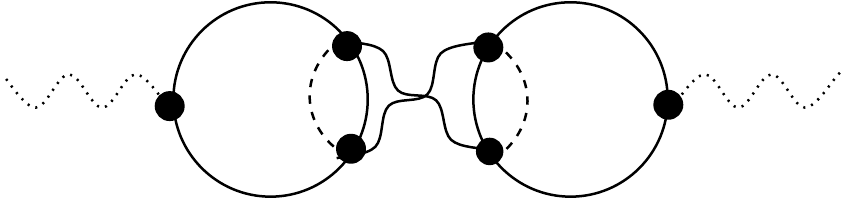}\,.
\end{equation}
In \cite{Patel:2022gdh}, MT graphs and AL graphs vanish due to the spatial delta, which will be illustrated in section \ref{sec:discussion}. In our case, although AL graphs are also zero for the same reason, MT diagrams \eqref{graph:mt} are non-trivial.
In $(2+1)$ dimensions, the contribution from MT diagrams reads
\begin{eqnarray}\label{eqn:mtcancel}
&&\tilde{\Pi}_{\mathrm{MT}}^{ab}(\bi\Omega_m)	\nn
 &=&\frac{1}{4m^4}\frac{1}{(2\pi)^8}\int d^2\bq d\omega_1d\omega_2 d^2\bk_1d^2\bk_2 \bk_1^a\bk_2^b (\bk_1+\bk_2)_{a'}(\bk_1+\bk_2)_{b'}  G(\bi \omega_1,\bk_1)G(\bi (\omega_1+\Omega_m),\bk_1)\nn
	&&\times G(\bi \omega_2,\bk_2)G(\bi (\omega_2+\Omega_m),\bk_2)\cdot D^{a'b'}(\bi(\omega_1-\omega_2),\bq)\nn
 &=&\delta^{ab}\frac{K^2v_F^4}{8}\frac{\calN^2}{(2\pi)^4} \int d^2\bq d\omega_1d\omega_2 d\xi_{\bk_1}d\xi_{\bk_2}
 \frac{1}{\bi\frac{\Gamma}{2}\sgn(\omega_1)-\xi_{\bk_1}}
 \frac{1}{\bi\frac{\Gamma}{2}\sgn(\omega_1+\Omega_m)-\xi_{\bk_1}}\nn
 &&\times \frac{1}{\bi\frac{\Gamma}{2}\sgn(\omega_2)-\xi_{\bk_2}}
 \frac{1}{\bi\frac{\Gamma}{2}\sgn(\omega_2+\Omega_m)-\xi_{\bk_2}}\frac{1}{\bq^2+c_0|\omega_1-\omega_2|}\nn
 &=&\frac{K^2v_F^4}{4}\frac{1}{2\pi} \frac{\calN^2}{\Gamma^2}\int q dq \int_{-\Omega_m}^0 d\omega_1\int_{-\Omega_m}^0 d\omega_2\frac{1}{\bq^2+c|\omega_1-\omega_2|}\nn
 &=&\frac{K^2v_F^4\calN^2}{16\pi\Gamma^2}\int_{-\Omega_m}^0 d\omega_1 d\omega_2\int_{-\Omega_m}^0\ln{\frac{\Lambda_q^2}{c|\omega_1-\omega_2|}}\nn
 &=&\frac{K^2v_F^4\calN^2}{32\pi\Gamma^2}\Omega_m^2\ln{\left(\frac{\me^3\Lambda_q^4}{c^2|\Omega_m|^2}\right)}.
\end{eqnarray}
Therefore the MT correction \eqref{eqn:mtcancel} precisely cancels the contribution from the electron self-energies \eqref{eqn:pi1}. This does not imply that the boson-electron scattering does not contribute to the resistivity. In fact, up to two loops, there is another polarization bubble to be considered, 
\begin{equation}\label{graph:new}
	\includegraphics[scale=0.35]{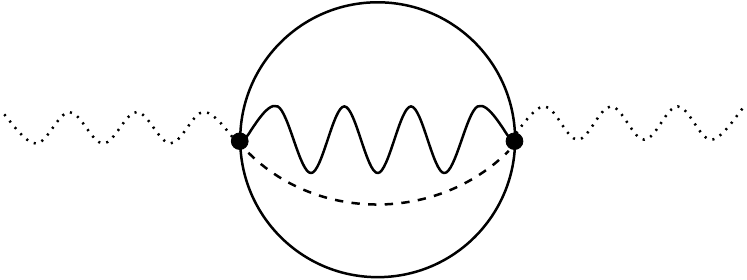}\,,
\end{equation}
which is obtained from $\mathscr{A}^{\mu}a_{\mu}\psi^{\dagger}\psi$, and this graph is crucial to produce the linearity.\\

Since the polarization \eqref{graph:new} is generated by $\mathscr{A}_{\mu}a^{\mu}\psi^{\dagger}\psi$, it does not exist in \cite{Patel:2022gdh}. In \cite{Kim:1994tfo}, eqn. \eqref{graph:new} is not considered because it yields a higher order-term in frequency. Since the boson propagator \eqref{eqn:b-propagator} has been dramatically changed, there is no reason to ignore this diagram. In fact, the polarization \eqref{graph:ggd} reads
\begin{eqnarray}\label{eqn:2loops}
    &&\tilde{\Pi}_{K(3)}^{ab}(\bi\Omega_m)\nn
    &=&\delta^{ab}\frac{K^2}{2m^2}\calN^2\int d\xi_{\bk_1}d\xi_{\bk_2} \frac{d^2\bq}{(2\pi)^2}\frac{d\omega_1}{2\pi}\frac{d\omega_2}{2\pi}
    \frac{1}{\bi\frac{\Gamma}{2}\sgn(\omega_1)-\xi_{\bk_1}}\frac{1}{\bi\frac{\Gamma}{2}\sgn(\omega_2)-\xi_{\bk_2}}
    \frac{1}{\bq^2+c_0|\Omega_m-\omega_1-\omega_2|}\nn
    &=&\delta^{ab}\frac{K^2}{16 m^2}\frac{\calN^2}{2\pi} \Omega_m ^2 \left(2 \log \left(\frac{\Lambda_q^4}{c_0^2\Omega_m^2 }\right)-2 \log \left(\frac{\Lambda^4}{c_0^2}\right)+\log (\Omega_m^2 )+3-2 \bi \pi \right).
\end{eqnarray}

Because we assumed that  $A_{\mu}a^{\mu}\psi_i^{\dagger}\psi_j$ is characterized by a coupling parameter $\check{K}_{ij}(\br)$  satisfying  $\langle\check{K}_{ij}(\br)\rangle=0$ and $\langle\check{K}_{ij}^*(\br)\check{K}_{i'j'}(\br')\rangle=K^2\delta_{ii'jj'}\delta(\br-\br')$,   
the diagrams such as
\begin{equation}\label{graph:other}
	\includegraphics[scale=0.35]{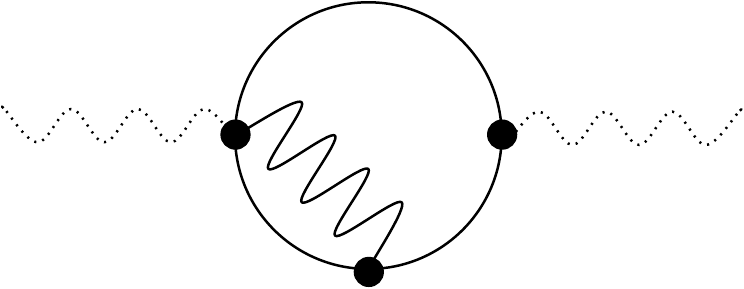}\,
\end{equation}
vanishes, and only diagrams  with an even number of vertices (of the same type) survive in our model. If we use only $K_{ij}$ without introducing $\tilde K_{ij}$, then this graph would be non-vanishing. Its contribution would read 
\begin{eqnarray}
	 &&\tilde{\Pi}_{K(4)}^{ab}(\bi\Omega_m)\nn
	 &=&\frac{K^2}{2m^3}\int\frac{d\omega_1}{2\pi}\frac{d\omega_2}{2\pi}\frac{d^2\bq}{(2\pi)^2}\frac{d^2\bk_1}{(2\pi)^2}\frac{d^2\bk_2}{(2\pi)^2}
	 \bk_1^a(\bk_1+\bk_2)^b
\nn
	  &&\times 	  \frac{1}{\bi\frac{\Gamma}{2}\sgn(\omega_1)-\xi_{\bk_1}} \frac{1}{\bi\frac{\Gamma}{2}\sgn(\omega_1+\Omega_m)-\xi_{\bk_1}}
	   \frac{1}{\bi\frac{\Gamma}{2}\sgn(\omega_2)-\xi_{\bk_2}} \frac{1}{\bq^2+c_0|\omega_1-\omega_2|}\nn
	  &=&\frac{K^2v_F^2}{32m(2\pi)}\int d\omega_1 d\omega_2 \sgn(\omega_2)
	  \frac{\sgn(\omega_1)-\sgn(\omega_1+\Omega_m)}{\bi\frac{\Gamma}{2}\sgn(\omega_1)-\bi\frac{\Gamma}{2}\sgn(\omega_1+\Omega_m)}
	  \ln\left(\frac{\Lambda_q^2}{c_0|\omega_1-\omega_2|}\right)\nn
	  &=&-\bi\frac{K^2v_F^2}{64m(2\pi)\Gamma}\Omega_m^2\ln\left(\frac{\me^3\Lambda_q^4}{c_0^2|\Omega_m|^2}\right).
\end{eqnarray}
This term cannot precisely yield linear-$T$ dependence, as it becomes a pure imaginary number after we apply Kubo formula. 
This is why we introduced $\check{K}_{ij}$ to get rid of the diagram \eqref{graph:other} for simplicity, which is of no consequence.
Moreover, two-loop corrections to fermion self-energy \eqref{graph:ddg} is of higher order in frequency. Hence, what we compute above is enough to capture the low-temperature feature of the model.\\

The conductivity $\sigma^{11} (\Omega)$ in \emph{real frequency} $\Omega$ can be obtained from the Kubo formula \cite{Kim:1994tfo,Coleman2019}
\begin{eqnarray}\label{eqn:kubo}
	\sigma^{\mu\nu}(\Omega)=-e^2\frac{\tilde{\Pi}^{\mu\nu}(\bi\Omega_m\to \Omega+\bi 0)}{\bi\Omega}.
\end{eqnarray}
Substituting the first-order polarization without loop corrections (\ref{eqn:gauge_e_se}) into the Kubo formula (\ref{eqn:kubo}), one obtains 
\begin{equation}\label{eqn:drude1}
	\sigma_0(\Omega)=\frac{n_e e^2}{m}\frac{1}{\Gamma+\bi\Omega}, 
\end{equation}
which is the usual Drude-like optical conductivity \cite{Coleman2019}.  Taking into account  the one-loop corrections to the polarisations from eqn.(\ref{eqn:2loops}), one expects the disorder scattering rate $\Gamma$ will be replaced by the \emph{transport scattering rate} $\tau_{\mathrm{tr}}$.
To find $\tau_{\mathrm{tr}}$, one can focus on the real part of the total conductivity or resistivity, as we are dealing with low frequencies. 
Substituting the polarisations (\ref{eqn:gauge_e_se}), (\ref{eqn:2loops}) into the conductivity (\ref{eqn:kubo}), and taking the real part, one obtains
\begin{eqnarray}\label{eqn:conductivity}
	\Re{\sigma(\Omega)}&=&\Re{\sigma_0(\Omega)+\sigma_{K}(\Omega)}\nn
	&=&e^2\frac{\calN v_F^2}{2\Gamma}-\frac{e^2K^2}{16 m^2}\frac{\calN^2}{2}\Omega,
\end{eqnarray}
at low frequencies.
It is straightforward to find the resistivity
\begin{eqnarray}\label{eqn:resistivity}
 \Re{\frac{1}{\sigma(\Omega)}}&=&
 \Re{\frac{1}{\sigma_0(\Omega)+\sigma_{K}(\Omega)}}\nn
 &\simeq&\Re{\frac{1}{\sigma_0(\Omega)}}-\Re{\frac{\sigma_{K}(\Omega)}{\sigma_0^2(\Omega)}}\nn
 &=&\frac{2}{e^2\calN v_F^2}\left(\Gamma+\frac{K^2\Gamma^2\calN}{16m^2v_F^2}|\Omega|\right)\nn
 &\equiv&\frac{m}{n_e e^2}\left(\Gamma+\frac{K^2\Gamma^2\calN}{16k_F^2}|\Omega|\right).
\end{eqnarray}
Therefore, one again obtains a linear-$T$ resistivity from an FS coupled to a bosonic vector field. \\

According to the calculation in this section, the spatial random disorder can yield linear-$T$ resistivity as well in the model considered in this paper. Comparing the resistivity (\ref{eqn:resistivity}) with the one in \cite{Patel:2022gdh}, one finds that they take similar forms, both satisfying $\rho=\rho_0+AT$, with $A$ proportional to the (averaged) square of the coupling parameter between fermions and bosons. However, as we have seen above, the linear resistivity in our model arises from $\mathscr{A}_{\mu}a^{\mu}\psi^{\dagger}\psi$ instead of $\ba\psi^{\dagger}\nabla\psi$, which is a fundamental difference from the one studied in \cite{Patel:2022gdh}.
 We could  expect that,  because the scaling dimension of the vector is one higher than that of the scalar, two different types of interaction could generate the same  temperature scaling. 
 In the following section, we generalize the system to $(3+1)$ dimensions 
 and illustrate that the transport properties are indeed independent of the boson type, such that a QED-like interaction and a Yukawa-like interaction will share the same qualitative properties.  However, we will find that the resistivity is no longer linear in temperature in higher spatial dimensions, as \emph{the scaling dimension of fields crucially depends on the   dimensionality of spacetime. }

 \paragraph{Difference between vector model and scalar model}
 Though the vector model in this article also yields a linear resistivity, the computation is rather different from the scalar (Yukawa) model in \cite{Patel:2022gdh}. In spatially random Yukawa theory, the linearity comes from the electron self-energy \eqref{graph:self1}. The vertex corrections including MT diagram and AL diagrams are zero due to the symmetry of the integrand. In contrast, our vector model is characterized by the vertex $\psi^{\dagger}\ba\nabla\psi$, each of which contains a factor of $(k_1+k_2)$. Therefore, graphs such as MT diagram \eqref{graph:mt} do not necessary vanish. Unlike the spatially random Yukawa theory \cite{Patel:2022gdh}, the vector model in this article can receive higher-order corrections to linearity. It has been illustrated that MT graph \eqref{graph:mt} precisely cancels the contribution from self-energy \eqref{graph:self1}. Instead, the key to linear-$T$ resistivity in a vector model is the coupling $\mathscr{A}^{\mu}a_{\mu}\psi^{\dagger}\psi$, graphically represented by \eqref{graph:ggd}. Consequently, the coefficient of $T$ takes a basically different form from the one in \cite{Patel:2022gdh}.
 
 \paragraph{Difference between vector models with and without spatial randomness}
 In addition to the fact that the model in this article couples the FS to a vector field instead of a regular $U(1)$ gauge field, there are other important differences between our model witha normal FS-gauge coupling model studied in \cite{Kim:1994tfo}. Firstly, because the coupling parameter has zero expectation value in this article, only diagrams with an even number of same vertices are non-vanishing. Therefore, the number of diagrams is reduced in our theory. More importantly, in \cite{Kim:1994tfo}, the coupling $\mathscr{A}_{\mu}a^{\mu}\psi^{\dagger}\psi$ results in a higher-order term, so it is neglected. One thus obtains a would-be $T^{4/3}$ resistivity from self-energies with the vertex $\psi^{\dagger}\nabla_a\psi\ba^a$. On the other hand, we find that the diagram \eqref{graph:ggd} is not higher-order once impurity and spatial randomness are imposed, which means that they should be taken into account. Due to the cancellation between the self-energy term \eqref{graph:self1} and the MT graph \eqref{graph:mt}, the only non-trivial contribution is from eqn.\eqref{graph:ggd}. In other words, the $T$-dependence  in our case is a result of $\mathscr{A}_{\mu}a^{\mu}\psi^{\dagger}\psi$, instead of $\psi^{\dagger}\nabla_a\psi\ba^a$.

\section{Breakdown of Linear-T Resistivity  in Higher Dimensions}\label{sec:3d}
 One of  our interest is to examine whether the mechanism of SYK-like randomization works in dimensions other than two, so we want to study the model in higher dimensions. Although  most of the  unconventional superconductors  and  their strange-metal regimes are observed in $(2+1)$ dimensional systems, yet there are examples of $3D$ superconductor showing linear-$T$ behavior in their normal phase
  \cite{Nguyen_2021}. Therefore, one might expect that the linearity is independent of dimensionality \cite{Legros_2018}. 
The calculation offers a non-trivial test of this conjecture  and  it will also determine whether the SYK-rization always bestows the linear-$T$ resistivity even in higher dimensions, helping us to understand the underlying mechanics of strange metals.\\

It turns out that the random disorder  with  spatial dependence loses its power in higher dimensions.
This  happens not only in the model  considered in this paper, but also in the model discussed in \cite{Patel:2022gdh}. However,  we will   find that these two systems  share the same temperature dependence in the transport, which supports the result in section \ref{sec:2d}.\\

\subsection{Three dimensional Fermi surface coupled with a vector field}
Generalized to $(3+1)$ dimensions, the vector field in the model (\ref{eqn:2daction}) has three spatial degrees of freedom. Let $a,b=1,2,3$, and one obtains a similar $G-\Sigma$ theory with saddle point equations yielding the propagators as well as self-energies,
\begin{eqnarray}
	&&G=G_*=\left(-\partial_\tau-\varepsilon_k+\mu-\Sigma\right)^{-1}\\
	&&\Sigma=\Sigma_*=
	v^2 G(\tau,\br)\delta^3(\br)
	+\frac{1}{m^2}\frac{(k_1+k_2)^a}{2}\frac{(k_1+k_2)^b}{2}D_{ab} G\bar{\delta}
	+\frac{1}{4m^2}D_{\mu\nu}D^{\mu\nu}G\bar{\delta}\\
	&&D_{ab}={D_*}_{ab}=K^2(-(-\partial_\tau^2+\bq^2)g^{ab}-K^2\Pi^{ab})^{-1}\\
	&&\Pi_{ab}={\Pi_*}_{ab}=-\frac{1}{m^2}\frac{(k_1+k_2)_a}{2}\frac{(k_1+k_2)_b}{2}G\bar{\delta}\cdot G
	-\frac{1}{4m^2}G D_{ab}G\bar{\delta},
\end{eqnarray}
where $a,b=1,2$. \\

Repeating the computation in section \ref{sec:2d}, one can first find the fermion self-energy
\begin{eqnarray}
	\Sigma_v(\bi\omega_n)&=&v^2\int\frac{d^3 \bk'}{(2\pi)^3}G(\bi\omega_n,\bk')\nonumber\\
	&=&v^2\int \mathcal{N} d\xi_{\bk'}\frac{1}{\bi\omega_n-\xi_{\bk'}-\Sigma(\bi\omega_n,\bk')}\nonumber\\
	&=&-\bi\frac{\Gamma}{2}\sgn(\omega_n),
\end{eqnarray}
where  $\Gamma\equiv2\pi v^2 \mathcal{N}$ is again the disorder scattering rate, and $\mathcal{N}=m^{3/2}\sqrt{\varepsilon_{\bm{k}_F}}/(2\pi^2)$ is the DoS at the Fermi energy in three dimensions. This will bring a constant residual resistivity, which is basically the same as a $(2+1)$-dimensional case in the presence of an FS. 

So, the dominant part of the polarization is
\begin{eqnarray}\label{eqn:gauge_e_se2}
	&&\tilde{\Pi}_0^{ab}(\bi\Omega_m)\nn
	&=&\frac{1}{m^2}\calN\int\frac{d\omega_n}{2\pi}\int d\xi_{\bk}
	\frac{k^ak^b}{\bi\frac{\Gamma}{2}\sgn(\omega_n)-\xi_{\bk}}
	\left(\frac{1}{\bi\frac{\Gamma}{2}\sgn(\omega_n+\Omega_m)-\xi_{\bk}}
	-\frac{1}{i\frac{\Gamma}{2}\sgn(\omega_n)-\xi_{\bk}}\right)\nn
	&=&\frac{v_F^2}{3}\delta^{ab}\calN \frac{1}{\Gamma} |\Omega_m|,
\end{eqnarray}
where we impose the isotropy such that the conductivity is diagonal. \\

Similarly, one obtains the boson self-energy
\begin{eqnarray}
	&&-K^2\Pi_0^{ab}(\bi\Omega_m)\nn
	&=&\frac{K^2}{6m^2}\calN^2\int\frac{d\omega_n}{2\pi}\int d\xi_{\bk}d\xi_{\bk'}
	\frac{k_F^2\delta^{ab}}{i\frac{\Gamma}{2}\sgn(\omega_n)-\xi_{\bk}}
	\left(\frac{1}{i\frac{\Gamma}{2}\sgn(\omega_n+\Omega_m)-\xi_{\bk'}}
	-\frac{1}{i\frac{\Gamma}{2}\sgn(\omega_n)-\xi_{\bk'}}\right)\nn
	&=&\pi\calN^2K^2\frac{v_F^2}{12}|\Omega_m|\delta^{ab}\nn
	&\equiv& c_0 \delta^{ab}|\Omega_m|.
\end{eqnarray}
The boson propagator takes the same form as the vector propagator in $(2+1)$ dimensions.\\

This will then give us the fermion self-energies from the electron-boson coupling,
\begin{eqnarray}
	\Sigma_K(\bi\omega_n)&=&
	\frac{K^2k_F^2}{4m^2}\calN\int\frac{d\Omega_m}{2\pi}\int\frac{d^3\bq}{(2\pi)^3}d\xi_{\bk}
	\frac{1}{i\frac{\Gamma}{2}\sgn(\omega_n+\Omega_m)-\xi_{\bk}}\frac{1}{\bq^2+c_0|\Omega_m|}\nn
	&=&-\bi\frac{K^2v_F^2}{4}\calN\int_{-\infty}^{+\infty}d\Omega_m\int_0^{\Lambda_q}\frac{4\pi |\bq|^2d|\bq|}{(2\pi)^3}
	\sgn(\omega_n+\Omega_m)\frac{1}{|\bq|^2+c_0|\Omega_m|}\nn
	&=&-\bi\frac{K^2v_F^2}{8\pi^2}\calN \int d\Omega_m\sgn(\omega_n+\Omega_m)\left(\Lambda_q-\frac{\sqrt{c_0|\Omega_m|}\pi}{2}\right)\nn
	&=& \bi\frac{K^2v_F^2}{12\pi}\calN \sqrt{c_0} |\omega_n|^{3/2},
\end{eqnarray}
where $\Lambda_q$ is a UV cut-off on $q$. One finds that the specific heat no longer takes a form of $\omega\ln(1/\omega)$, and the linear-$T$ resistivity is not expected to exist any more. To be more explicit, let us continue to the polarization at the next order, which reads
\begin{eqnarray}\label{eqn:qed3d}
	&&2\tilde{\Pi}_{K(1)}^{ab}(\bi\Omega_m)\nn
	&=&
	2\frac{1}{m^2}\calN\int\frac{d\omega_n}{2\pi}\int d\xi_{\bk} k^{a}k^{b}
	\left(\frac{1}{\bi\frac{\Gamma}{2}\sgn(\omega_n)-\xi_{\bk}}\right)^2\frac{1}{\bi\frac{\Gamma}{2}\sgn(\omega_n+\Omega_m)-\xi_{\bk}}\Sigma_K(\bi\omega_n)\Big|_{\Omega_m=0}^{\Omega_m}\nn
	&=&\bi\frac{K^2}{3}v_F^2\delta^{ab}\mathcal{N}\int d\omega_n
	\frac{-\sgn(\omega_n+\Omega_m)+\sgn(\omega_n)}{\left(\bi\frac{\Gamma}{2}\sgn(\omega_n+\Omega_m)-\bi\frac{\Gamma}{2}\sgn(\omega_n)\right)^2}
	\left[ \bi\frac{v_F^2}{12\pi}\calN \sqrt{c_0} |\omega_n|^{3/2}\right]\nn
	&=&\frac{1}{90\pi}K^2v_F^4\delta^{ab}\frac{\calN^2}{\Gamma^2}\sqrt{c_0}|\Omega_m|^{5/2}.
\end{eqnarray}
The vertex correction from MT diagrams reads
\begin{eqnarray}
&&\tilde{\Pi}_{\mathrm{MT}}^{ab}(\bi\Omega_m)	\nn
 &=&\frac{1}{6m^4}\frac{1}{(2\pi)^{11}}\int d^3\bq d\omega_1d\omega_2 d^3\bk_1d^3\bk_2 (\bk_1)^a(\bk_2)^b (\bk_1+\bk_2)_{a'}(\bk_1+\bk_2)_{b'}  G(\bi \omega_1,\bk_1)G(\bi (\omega_1+\Omega_m),\bk_1)\nn
	&&\times G(\bi \omega_2,\bk_2)G(\bi (\omega_2+\Omega_m),\bk_2)\cdot D^{a'b'}(\bi(\omega_1-\omega_2),\bq)\nn
 &=&\frac{K^2v_F^4}{12}\frac{\calN^2}{(2\pi)^5} \int d^3\bq d\omega_1d\omega_2 d\xi_{\bk_1}d\xi_{\bk_2}
 \frac{1}{\bi\frac{\Gamma}{2}\sgn(\omega_1)-\xi_{\bk_1}}
 \frac{1}{\bi\frac{\Gamma}{2}\sgn(\omega_1+\Omega_m)-\xi_{\bk_1}}\nn
 &&\times \frac{1}{\bi\frac{\Gamma}{2}\sgn(\omega_2)-\xi_{\bk_2}}
 \frac{1}{\bi\frac{\Gamma}{2}\sgn(\omega_2+\Omega_m)-\xi_{\bk_2}}\frac{1}{\bq^2+c_0|\omega_1-\omega_2|}\nn
 &=&-\delta^{ab}\frac{K^2v_F^4\calN^2}{90\pi\Gamma^2}\Omega_m^2\left(\Lambda_q-\sqrt{c_0}|\Omega|^{3/2}\right),
\end{eqnarray}
canceling again the self-energy contribution \eqref{eqn:qed3d} after taking the real part of the conductivity. The non-trivial temperature dependence thus comes from the two-loop polarization \eqref{graph:ggd}, which reads
\begin{eqnarray}\label{eqn:2loops2}
    &&\tilde{\Pi}_{K(3)}^{ab}(\bi\Omega_m)\nn
    &=&\delta^{ab}\frac{K^2}{4m^2}\calN^2\int d\xi_{\bk_1}d\xi_{\bk_2} \frac{d^3\bq}{(2\pi)^3}\frac{d\omega_1}{2\pi}\frac{d\omega_2}{2\pi}
    \frac{1}{\bi\frac{\Gamma}{2}\sgn(\omega_1)-\xi_{\bk_1}}\frac{1}{\bi\frac{\Gamma}{2}\sgn(\omega_2)-\xi_{\bk_2}}
    \frac{1}{\bq^2+c_0|\Omega_m-\omega_1-\omega_2|}\nn
    &=&-\delta^{ab}\frac{K^2}{16 m^2}\frac{\calN^2}{2\pi^2}\int d\omega_1 d\omega_2
    \sgn(\omega_1)\sgn(\omega_2)\left(\Lambda_q-\frac{\pi\sqrt{c_0}}{2}\sqrt{|\Omega-\omega_1-\omega_2|}\right)\nn
    &=&\delta^{ab}\frac{K^2}{16 m^2}\frac{\calN^2}{2\pi^2} \Omega_m ^2 \left(\frac{\Lambda_q}{2}-\frac{2}{15}\sqrt{c_0}\pi|\Omega_m|^{1/2}\right).
\end{eqnarray}
Similarly, if we take $\check{K}_{ij}=K_{ij}$, the graph \eqref{graph:other} will yield a polarization 
\begin{eqnarray}
	 &&\tilde{\Pi}_{K(4)}^{ab}(\bi\Omega_m)\nn
	&=&\frac{K^2}{2m^3}\int\frac{d\omega_1}{2\pi}\frac{d\omega_2}{2\pi}\frac{d^3\bq}{(2\pi)^3}\frac{d^3\bk_1}{(2\pi)^3}\frac{d^3\bk_2}{(2\pi)^3}
	(\bk_1)^a(\bk_1+\bk_2)^b
	\frac{1}{\bi\frac{\Gamma}{2}\sgn(\omega_1)-\xi_{\bk_1}} \frac{1}{\bi\frac{\Gamma}{2}\sgn(\omega_1+\Omega_m)-\xi_{\bk_1}}\nn
	&&\times
	\frac{1}{\bi\frac{\Gamma}{2}\sgn(\omega_2)-\xi_{\bk_2}} \frac{1}{\bq^2+c_0|\omega_1-\omega_2|}\nn
	&=&\frac{\delta^{ab}K^2v_F^2}{48 m\pi^2}\int d\omega_1 d\omega_2 \sgn(\omega_2)
	\frac{\sgn(\omega_1)-\sgn(\omega_1+\Omega_m)}{\bi\frac{\Gamma}{2}\sgn(\omega_1)-\bi\frac{\Gamma}{2}\sgn(\omega_1+\Omega_m)}
	\left(\Lambda_q-\sqrt{c_0|\omega_1-\omega_2|}\frac{\pi}{2}\right)\nn
	&=&-\bi\delta^{ab}\frac{K^2v_F^2}{96\pi\Gamma}\Omega_m^2\frac{2}{5}\Omega_m^{5/2}.
\end{eqnarray}
Using Kubo formula (\ref{eqn:kubo}), one finds
\begin{eqnarray}
	\Re{\sigma^{ab}_{K}}\propto\delta^{ab}\Omega_m^{3/2}.
\end{eqnarray}
Therefore, there is no linear-$T$ resistivity after the model (\ref{eqn:action3}) is generalized to higher dimensions. Similarly, we find that the spatially random Yukawa model in \cite{Patel:2022gdh} also loses linear-$T$ resistivity in $(3+1)$ dimensions.

\subsection{Random Yukawa coupling model in $(3+1)$ dimensions}
This subsection provides a straightforward generalization of the random Yukawa model of ref. \cite{Patel:2022gdh} and illustrates the nonlinear-$T$ resistivity. As the spatially uniform theory (described by $g$-coupling in \cite{Patel:2022gdh}) has no contribution to the resistivity due to the cancellation from vertex corrections \cite{Patel:2022gdh}, we will only consider the potential disorder and interaction disorder. For simplicity, let us relabel the parameter $g'$ in \cite{Patel:2022gdh} by $g$. Therefore, the action is
\begin{align}
	\mathcal{S}_g &= \int d\tau\sum_{\bk}\sum_{i=1}^N\psi^\dagger_{i\bk}(\tau)\left[\partial_\tau +\varepsilon (\bk) \right]\psi_{i\bk}(\tau)\nn
	&+\frac{1}{2}\int d\tau \sum_{\bq}\sum_{i=1}^N \phi_{i\bq}(\tau)\left[-\partial_\tau^2 + \bq^2 +m_b^2\right]\phi_{i,-\bq}(\tau)\nn
	&+\frac{1}{N} \int d\tau d^3 r \sum_{i,j,l=1}^N g_{ijl}(\br) \psi^\dagger_{i}(\br,\tau)\psi_{j}(\br,\tau)\phi_{l}(\br,\tau)\nn
	&+ \frac{1}{\sqrt{N}} \int d^3 r d \tau \, v_{ij} (\br)  \psi_i^{\dagger} (\br, \tau) \psi_j(\br, \tau),
	\label{eq:latticeaction}
\end{align} 
where the random coupling $g_{ijk}$ and $v_{ij}$ has zero mean and
\begin{eqnarray}
	&\langle v^*_{ij}(\br)v_{i'j'}(\br')\rangle&=v^2\delta_{ii'}\delta_{jj'}\delta(\br-\br'),\nn
	&\langle g^*_{ijl}(\br)g_{i'j'l'}(\br')\rangle&=g^2\delta_{ii'}\delta_{jj'}\delta_{ll'}\delta(\br-\br').
\end{eqnarray}
After performing the disorder average and introducing self-energies as dynamical degrees of freedom, one obtains the saddle point equations
\begin{eqnarray}\label{fullsaddle}
	&&\Sigma(\tau,\mathbf{r})=
	v^2G(\tau,\mathbf{r}=0)\delta^2(\mathbf{r})+{g}^2G(\tau,\mathbf{r}=0)D(\tau,\mathbf{r}=0)\delta^2(\mathbf{r})\nn
	&&\Pi(\tau,\mathbf{r})=-{g}^2G(-\tau,\mathbf{r}=0)G(\tau,\mathbf{r}=0)\delta^2(\mathbf{r})\nn
	&&G(i\omega_n,\bk)=\frac{1}{\bi\omega_n-\varepsilon_{\bk}+\mu-\Sigma(i\omega_n,\bk)}\nn
	&&D(i\Omega_m,\bq)=\frac{1}{\Omega_m^2+\bq^2+m_b^2-\Pi(i\Omega_m,\bq)}.
\end{eqnarray}
These take the same form as those from $(2+1)$ dimensions \cite{Patel:2022gdh}.\\ 

Similarly, one calculates first the self-energy from the potential disorder, 
\begin{eqnarray}\label{}
	\Sigma_{v}(\bi\omega_n,\bk)
	&=&v^2\int \calN d\xi_{\bq}\frac{1}{\bi\omega_n-\xi_{\bq}-\Sigma(i\omega_n,\bq)}\nn
	&=&-\bi\frac{\Gamma}{2}\sgn(\omega_n),
\end{eqnarray}
where  $\Gamma=2\pi v^2 \mathcal{N}$ and $\calN$ is the DoS at FS in three dimensions. Since $\Gamma$ is the largest scale in this system, the Fermion propagator can be approximated into
\begin{equation}
	G(\bi\omega_n,\bk)\simeq\frac{1}{\bi\frac{\Gamma}{2}\sgn(\omega_n)-\xi_{\bk}}
\end{equation}
at low frequencies.\\

Then, continuing to the boson self-energies, we obtain
\begin{eqnarray}
	&&\Pi_{g}(\bi\Omega_m)\nn
	&=&-{g}^2\calN^2\int\frac{d\omega_n}{2\pi}\int d\xi_{\bk}d\xi_{\bk'}
	\frac{1}{\bi\frac{\Gamma}{2}\sgn(\omega_n)-\xi_{\bk}}
	\left(\frac{1}{\bi\frac{\Gamma}{2}\sgn(\omega_n+\Omega_m)-\xi_{\bk'}}
	-\frac{1}{\bi\frac{\Gamma}{2}\sgn(\omega_n)-\xi_{\bk'}}\right)\nn
	&=&\frac{\bi}{2}{g}^2\calN^2\int d\omega_n\int d\xi_{\bk'}\sgn(\omega_n)
	\left(\frac{1}{\bi\frac{\Gamma}{2}\sgn(\omega_n+\Omega_m)-\xi_{\bk'}}
	-\frac{1}{\bi\frac{\Gamma}{2}\sgn(\omega_n)-\xi_{\bk'}}\right)\nn
	&=&\frac{\pi}{2}{g}^2\calN^2	\int d\omega_n\sgn(\omega_n)\left(\sgn(\omega_n+\Omega_m)-\sgn(\omega_n)\right)\nn
	&=&-\pi\calN^2{g}^2|\Omega_m|\nn
	&\equiv&-c_d|\Omega_m|.
\end{eqnarray}

One can use the result above to find the self-energy from the random Yukawa coupling, which reads
\begin{eqnarray}
	\Sigma_{g}(\bi\omega_n)&=&{g}^2\calN\int_{-\infty}^{+\infty}\frac{d\Omega_m}{2\pi}\int d\xi_{\bk}\int_0^{+\infty}\frac{q^2dq}{2\pi^2}
	\frac{1}{\bi\frac{\Gamma}{2}\sgn(\omega_n+\Omega_m)-\xi_{\bk}}\frac{1}{q^2+c_d|\Omega_m|}\nn
	&=&-\bi g^2\calN\int_{-\infty}^{+\infty}d\Omega_m\int_0^{+\infty}\frac{q^2dq}{4\pi^2}
	\sgn(\omega_n+\Omega_m)\frac{1}{q^2+c_d|\Omega_m|}\nn
	&=&\frac{-\bi{g}^2}{4\pi^2}\calN\int_{-\infty}^{+\infty}d\Omega_m
	\left(\Lambda_q-\frac{\pi}{2}\sqrt{c_d|\Omega_m|}\right)\sgn(\omega_n+\Omega_m)\nn
	&=&\frac{\bi{g}^2}{6\pi}\calN|\omega_n|^{3/2}.
\end{eqnarray}
We find that $T\ln(1/T)$ dependence no longer exists either, and there should be no linear-$T$ resistivity. To be explicit, let us compute the conductivity and see how the dimension changes the result. The derivation of conductivity in \cite{Guo2022} is still valid, so one can directly move on to the computation.\\

First is the first-order current-current correlation dominated by potential disorder,
\begin{eqnarray}
	\frac{1}{N}\tilde{\Pi}_v^{xx}(\bi\Omega_m)&=&
	v_F^2\calN\int_{-\pi}^{\pi}d\theta\cos[2](\theta)\int\frac{d\omega_n}{2\pi}
	\int d\xi_{\bk}\frac{1}{\bi\frac{\Gamma}{2}\sgn(\omega_n)-\xi_{\bk}}
	\frac{1}{\bi\frac{\Gamma}{2}\sgn(\omega_n+\Omega_m)-\xi_{\bk}}\nn
	&=&\frac{\calN v_F^2}{2\Gamma}\Omega_m.
\end{eqnarray}
So the conductivity is
\begin{equation}
	\frac{1}{N}\Re[\sigma_{\Sigma,v}]=-\frac{1}{N}\frac{\Im\left[\Pi_{v}^{xx}(\bi\Omega_m)-\Pi_{v}^{xx}(0)|_{\bi\Omega_m\to \Omega+i0^+}\right]}{\Omega}=\frac{\calN v_F^2}{2\Gamma},
\end{equation}
which is a constant, qualitatively the same with $(2+1)$ dimensional model in \cite{Patel:2022gdh}.\\

Then the contribution from $g$ term is
\begin{eqnarray}\label{eqn:yukawa3d}
	&&\frac{1}{N}\tilde{\Pi}_{g}^{xx}(\bi\Omega_m)\nn
	&=&
	2v_F^2\calN\int_{-\infty}^{+\infty}\frac{d\omega_n'}{2\pi}\int_{-\pi}^{+\pi}\frac{d\theta\cos^2\theta}{2\pi}
	\int_{-\infty}^{\infty}d\xi_{\bk}
	\frac{1}{\bi\frac{\Gamma}{2}\sgn(\Omega_m+\omega'_n)-\xi_{\bk}}
	\frac{1}{(\bi\frac{\Gamma}{2}\sgn(\omega_n')-\xi_{\bk})^2}\Sigma_{g}(\bi\Omega_m)\nn
	&=&\frac{\bi}{2}v_F^2\calN\int_{-\infty}^{+\infty}d\omega'_n
	\frac{-\sgn(\omega_n'+\Omega_m)+\sgn(\omega_n')}{(\bi\frac{\Gamma}{2}\sgn(\Omega_m+\omega_n')-\bi\frac{\Gamma}{2}\sgn(\omega_n'))^2}
	\left(\frac{\bi{g}^2}{6\pi}\calN|\omega_n|^{3/2}\right)\nn
	&=&-\frac{v_F^2}{30\pi}\frac{\calN^2{g}^2}{\Gamma^2}|\Omega_m|^{5/2}.
\end{eqnarray}
Similarly, the analytic continuation to the real axis bring us a conductivity $\propto \Omega^{3/2}$.\\ 

It is not surprising that the linear-$T$ resistivity from spatial random coupling cannot survive in higher dimensions. Let us close this section with a rough analysis and take the scalar model in \cite{Patel:2022gdh} for example.\\

As the vertex corrections vanish, the non-trivial $\Omega$- (or $T$-) dependence comes from the one-loop corrections to the polarization
\begin{equation}
	\includegraphics[scale=0.45]{eself-energy}\,.
\end{equation}
The qualitative behavior can be conveniently read from the electron self-energy
\begin{equation}
	\includegraphics[scale=0.3]{dg-self-energy}\,.
\end{equation}
Because the momentum conservation is removed on each vertex, the dependence on frequency or temperature arises from the integral 
\begin{eqnarray}\label{eq:t-depend}
	\int d\Omega_m  \int d^d\bq \int d\xi_{\bk} \frac{1}{\bi\frac{\Gamma}{2}\sgn(\omega_n+\Omega_m)-\xi_{\bk}}\frac{1}{\bq^2+c|\Omega_m|}
\end{eqnarray}
in $(d+1)$ dimensions with $c$ a constant.   
In eqn.\eqref{eq:t-depend}, $d\Omega_m$  will produce a term linear in frequency,  while $\int d\xi_{\bk}$ yields a  term at order ${\cal O}(\omega^0)$. The integral over $\bq$  in general will bring a term with dimension of  $\omega^{(d-2)/2 }$. As a consequence,  only when  $d=2$,  can these models bring us a linear-$T$ dependence.\\

Similarly,  the non-trivial $T$-dependence in the vector model arises from the polarization \eqref{graph:new}, which is proportional to
\begin{eqnarray}
	&&\int d\omega_1 d\omega_2 \int d^d\bq \int d\xi_{\bk_1} d\xi_{\bk_2} \frac{1}{\bi\frac{\Gamma}{2}\sgn(\omega_1)-\xi_{\bk_1}}\frac{1}{\bi\frac{\Gamma}{2}\sgn(\omega_2)-\xi_{\bk_2}}\frac{1}{\bq^2+c|\Omega_m-\omega_1-\omega_2|}\nn
	&\sim&\Omega_m^{(d+2)/2}.
\end{eqnarray}
The Kubo formula then yields a resistivity $\sim T^{d/2}$, so only when $d=2$ can we have linearity.

Therefore, the idea of using a spatially random Yukawa-type interaction to obtain a linear resistivity in \cite{Esterlis2021,Guo2022,Patel:2022gdh} works only in $(2+1)$ dimensions \cite{Sachdev2023}. 
Comparing the result of the scalar boson (\ref{eqn:yukawa3d}) with that of the vector boson (\ref{eqn:qed3d}), one finds that they take a similar form, implying that the effects of spatial random coupling to the resistivity is independent of the type of bosons.\\

\section{The Emergence of Linearity: Fermi Liquid vs Strange Metal }\label{sec:discussion}

So far, we have shown that boson-electron scatterings can be the source of linear resistivity, in spite of the boson types. To show the linearity, the ref. \cite{Patel:2022gdh} and this article performed extensive calculations. 
We found that different Feynman diagrams are used to get the same qualitative property. 
 Our purpose here is that the qualitative properties can be read directly without cumbersome computation. 
 We will follow Bloch's argument \cite{Bloch1929,bloch1930elektrischen,Ashcroft_Mermin_1976}  to give an intuitive understanding of this linear resistivity and its dimensional dependence.\\

First, the scattering rate contains contribution from DoS and a small-angle correction $(1-\cos\theta)$, with $\theta$ the scattering angle,
such that the total transport  rate is
\begin{equation}\label{eqn:doscos}
	\rho\sim \frac{1}{\tau_{\mathrm{tr}}}\sim  \int \mathrm{DoS}\cdot(1-\cos\theta).
\end{equation}
 We will use eqn.\eqref{eqn:doscos} to see which models have linear-$T$ dependence and why this dependence is difficult to obtain.\\

\begin{figure}[htbp]
	\centering
	\includegraphics[width=.25\textwidth]{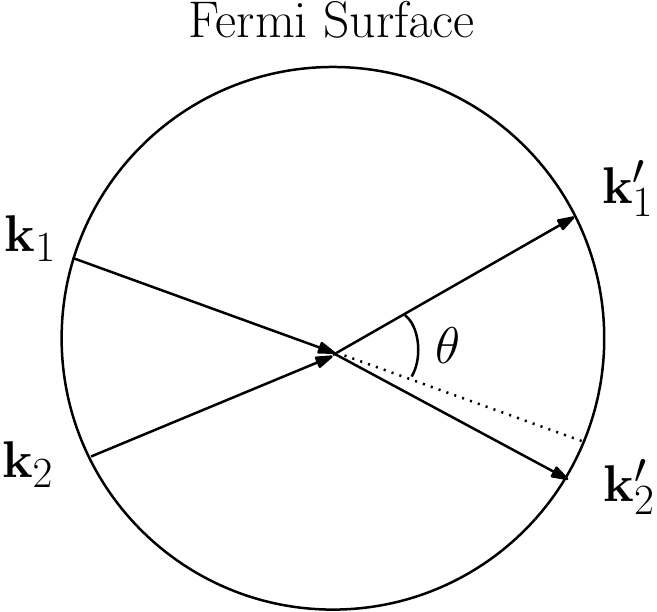}
	\caption{An illustration of a binary quasiparticle collision. \label{fig:i}}
\end{figure}

\paragraph{Fermi Liquid ($\sim T^2$)} Let us start with the Fermi liquid theory, which is characterized by the $  T^2$ resistivity. 
The quadratic behavior originates from collisions among quasiparticles. The simplest model is the binary collision   by   the $4$-fermion interaction term,  $u\psi^{\dagger}\psi^{\dagger}\psi\psi$ \cite{Coleman2019,Sachdev_2011}.
 Two particles with momentum $\bk_1$ and $\bk_2$ scatter into $\bk_1'$ and $\bk_2'$,
\begin{eqnarray}
	\bk_1+\bk_2\to \bk_1'+\bk_2',
\end{eqnarray} 
as is shown in Fig. \ref{fig:i}. 
\\
This process can also be viewed as the decay of one quasi-particle into two quasi-particles and one quasi-hole, so the scattering rate can be calculated as the decay rate of particle ($\bk_1$)  via Fermi's golden rule 
  \cite{Coleman2019,Mahan2014}: 
  
\begin{eqnarray}
	\rho&\sim& \int d\bk_2d\bk_1'd\bk_2'\delta(\bk_1+\bk_2-\bk_1'-\bk_2')\delta(\epsilon_1+\epsilon_2-\epsilon_1'-\epsilon_2')
	(1-\cos\theta)f_{\bk_2}(1-f_{\bk_1'})(1-f_{\bk_2'})\nn
	&=&\int d\bk_1'd\bk_2'	\delta(\epsilon_1+\epsilon_2-\epsilon_1'-\epsilon_2')(1-\cos\theta)f_{\bk_1'+\bk_2'-\bk_1}(1-f_{\bk_1'})(1-f_{\bk_2'})\nn
	&\propto&\int d\epsilon_1'd\epsilon_2'f(-\epsilon_1')f(-\epsilon_2')f(\epsilon_1'+\epsilon_2'-\epsilon_1)\nn
	&\sim&T^2\int dx dy\frac{1}{1-e^{-x}}\frac{1}{1-e^{-y}}\frac{1}{1-e^{z}},
\end{eqnarray}
where $f_{\bk}$ is the Fermi function. 
The essential  steps to the $T^2$ are as follows.
\begin{itemize}
	\item[Step 1.] The momentum conservation reduces one momentum integral. The energy conservation, along with the distribution functions, restricts the integration to range around the FS. 
	\item[Step 2.] We split momentum integral into radial and angular directions, $d^d\bk \sim D_d(\epsilon)d\epsilon d\Omega$, where
	$D_d(\epsilon)=\epsilon^{(d-2)/2} \sim \epsilon_F^{(d-2)/2} \sim $ constant  due to the FS restriction. 
	\item[Step 3.] The angular integral does not contribute to the temperature scaling, so the momentum integration  is reduced to the energy integral.
	\item[Step 4.] Take $x\equiv\epsilon_1'\beta$, $y\equiv\epsilon_2'\beta$, and $z\equiv\epsilon_2\beta$, with $\beta\equiv1/(k_BT)$.
\end{itemize}

Therefore, the $T^2$-dependent resistivity comes merely from the DoS of fermions in Fermi liquids. 
In all these steps, the presence of the FS is essential, and  this result holds true in all dimensions.
 Notice, however, that to get the constant piece of the resistivity,  this $4$-electron interaction model should also introduce the random disorder potential term. 
\\

\paragraph{Yukawa Coupling ($\sim T^2$)} A similar quadratic resistivity can also be generated from (random) electron-boson scatterings, $g_{ijl}\psi^{\dagger}_i\psi_j\phi_l$ \cite{Guo2022},  and  a spatial  disorder potential,  $v_{ij}(\br)\psi^{\dagger}_{i}\psi_j$ \footnote{Appendix \ref{App:summary} shows that the flavor disorder hardly changes the qualitative properties.}. In most cases, in addition to the electron self-energy contributions \eqref{graph:self1}, 
 we also have to  include the vertex corrections given by the  MT diagram.

Although the self-energy \eqref{graph:self1} 
brings a resistivity linear in $T$   \cite{Guo2022,Patel:2022gdh}, this linear dependence is canceled by
 the MT graph \eqref{graph:mt}. Therefore, one has to go to the next order, i.e. AL diagrams \eqref{graph:al1}, which 
 result in  a  resistivity $\rho\sim T^2$.  
 Since it does not involve the spatially random Yukawa coupling, it can be considered as a model for the Fermi liquid.
 A limitation of this model is that it works only  in $(2+1)$ dimension.\\

\paragraph{Strange metal from the Spatially Random Yukawa Coupling} 
 On the other hand, the spatially random scalar-fermion model  in \cite{Patel:2022gdh} receives no vertex corrections, so the term linear in $T$ survives.
 To see the reason, let us take the MT diagram (\ref{graph:mt}) as an example. Usually its contribution reads
\begin{eqnarray}\label{eqn:vertexc}
	&&\int d\omega_1d\omega_2 d\bk_1d\bk_2 \bk_1\bk_2 G(\bi \omega_1,\bk_1)G(\bi (\omega_1+\Omega_m),\bk_1)\nn
	&&\times G(\bi \omega_2,\bk_2)G(\bi (\omega_2+\Omega_m),\bk_2)D(\bi(\omega_1-\omega_2),\bk_1-\bk_2).
\end{eqnarray}
 Notice that the delta function correlation of the spatial random coupling  implies that the vertices $w$ and $w'$ in \eqref{graph:mt} should be contracted. 
After the contraction of vertex $w$ and $w'$ in a electron self-energy diagram, the boson propagator becomes a loop, as illustrated below,
\begin{equation}
	\includegraphics[scale=0.5]{dg-self-energy-loop}\,.
\end{equation}
This means that the momentum conservation imposed on the vertices is relaxed and the bosonic momentum $\bq$ is decoupled from the fermion momentum $\bk$ so that 
  the integral (\ref{eqn:vertexc}) becomes 
\begin{eqnarray}\label{eqn:mt}
	&&\int d\omega_1d\omega_2 d\bk_1d\bk_2 d\bq \bk_1\bk_2  G(\bi \omega_1,\bk_1)G(\bi (\omega_1+\Omega_m),\bk_1)\nn
	&&\times G(\bi \omega_2,\bk_2)G(\bi (\omega_2+\Omega_m),\bk_2)\cdot D(\bi(\omega_1-\omega_2),\bq). 
\end{eqnarray}
 Now notice that the integrand is an odd function of $\bk_1$ and $\bk_2$ since $G$'s are even functions,
  so the MT diagram vanishes \cite{Patel:2022gdh}. All the AL diagrams vanish for the same reason. As a consequence,  the self-energy contribution given by  \eqref{graph:self1}  survives and  the resistivity is linear-$T$.  \\

\paragraph{Strange metal from Spatially Random Vector Field Coupling}
In this paper, we consider an FS coupling to a vector field. In contrast to the Yukawa vertex $\psi^{\dagger}\psi\phi$, our vertex $\psi^{\dagger}\nabla\psi\ba$ receives an extra factor $(\bk_1+\bk_2)/2$. Consequently, the MT diagram \eqref{eqn:mt} becomes 
\begin{eqnarray}\label{eqn:mt2}
	&&\int d\omega_1d\omega_2 d\bk_1d\bk_2 d\bq \bk_1\bk_2 (\bk_1+\bk_2)_{\mu}(\bk_1+\bk_2)_{\nu}  G(\bi \omega_1,\bk_1)G(\bi (\omega_1+\Omega_m),\bk_1)\nn
	&&\times G(\bi \omega_2,\bk_2)G(\bi (\omega_2+\Omega_m),\bk_2)\cdot D(\bi(\omega_1-\omega_2),\bq),
\end{eqnarray}
which is an even function of $\bk_i$. In other words, the MT diagram is not zero in our model. As is shown in Section \ref{sec:2d} and \ref{sec:3d}, the MT diagram precisely cancels the self-energy contribution to the polarization. To obtain linear-$T$ resistivity, we have to consider the two-loop diagram \eqref{graph:ggd}. Though sharing the similar qualitative behavior, the linearity of random Yukawa model arises from $\mathscr{A} \psi^{\dagger}\nabla\psi$, whereas that of our vector model is from $\mathscr{A} \ba \psi^{\dagger}\psi$. This means that for the vector model, the polarization bubble of the external field will also be contracted due to randomization.

\paragraph{Decay-rate argument} 
The strange metal analysis above is from the perspective of quantum field theory. Alternatively, one can also apply eqn.\eqref{eqn:doscos} to 
find the  dependence on temperature.   
As is shown in Fig.\ref{fig:ii} below, three particles are involved in this process. A boson with wave-vector $\bq$ scatters an electron with momentum $\bk_1 $ to the one with $\bk_2 $, and the scattering angle $\theta$ is very small at low temperatures. In this case, the source of resistivity is the boson, so the DoS is evaluated over the boson wave-vector $\bq$, unlike the $4$-fermion interaction model of Fermi liquid. 

\begin{figure}[htbp]
	\centering
	\includegraphics[width=.2\textwidth]{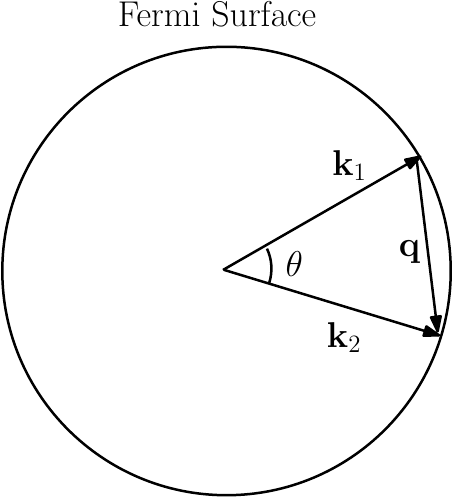}
	\caption{An illustration of an electron-boson scattering at low temperatures. \label{fig:ii}}
\end{figure}

Consider a theory in $d$ spatial dimensions with bosons whose dispersion relation is  $\epsilon\sim \bq^\alpha$. Then contributions from the DoS of bosons to the decay rate
are
\begin{equation}\label{eqn:dos}
	\int d^d\bq\frac{1}{e^{\beta\epsilon}-1}\sim T^{d/\alpha}\int dx\frac{x^{(d-\alpha)/\alpha}}{e^x-1}. 
\end{equation}
In the last step,  we set $\beta\epsilon =x$ and we used the fact that for low temperature case ($\beta\sim \infty$), the integration limit  of $x$  is approximately  from $0$ to $\infty$. 
Therefore, the scattering rate contains a factor of $T^{d/\alpha}$ from DoS.  
On the other hand, the angular dependence here contributes to the temperature power 
(unless the model has a spatially random coupling), 
where
\begin{equation}\label{eqn:scatter}
	(1-\cos\theta)\simeq  \frac{\bq^2}{k_F^2}\sim \epsilon^{2/\alpha}/{k_F^2}\sim T^{2/\alpha} x^{2/\alpha}/{k_F^2}. 
\end{equation}
Substituting eqns.\eqref{eqn:dos} and \eqref{eqn:scatter} into eqn.\eqref{eqn:doscos}, one obtaines the resistivity $\sim T^{(d+2)/\alpha} $.\\

In the spatially uniform model \cite{Esterlis2021,Guo2022,Patel:2022gdh}, our model, and the spatially random Yukawa theory \cite{Patel:2022gdh}, the boson self-energies are all approximately linear in frequency, taking similar form to eqns. (\ref{eqn:seboson}). The boson propagator then reads
\begin{equation}\label{eqn:b-propagatorb}
	D(\bi\Omega_m,\bq)\simeq\frac{1}{\bq^2+c_0|\Omega_m|},
\end{equation}
so poles are along $\epsilon \sim \bq^2$. Therefore, $\alpha=2$, and the contribution from DoS of bosons is  $T^{d/2}$ according to eqn.\eqref{eqn:dos}. 
Regarding the contribution from $(1-\cos\theta)$, two cases are different. 
\begin{enumerate}
	\item  For the model with a spatially uniform coupling (with or without flavor randomness), 
	$(1-\cos\theta)\sim \theta^2\sim \bq^2\sim \epsilon\sim T$. Taking $d=2$, the total transport scattering rate is proportional to $T^2$, which matches the result in \cite{Guo2022}. 
	Unlike the resistivity of Fermi liquids which is $T^2$ in all dimensions,  the quadratic resistivity from (random) Yukawa coupling with random potential only appears in $(2+1)$ dimensions. 
	\item On the other hand,  for the models with  spatially random  couplings, the delta function in the correlator of couplings  removes the contribution  from $(1-\cos\theta)$ \footnote{The  contribution from  $(1-\cos\theta)$ is obtained only when momentum conservation is imposed.}.
			In such models,  bosons can always relax the electric current even at very  low temperatures. Consequently, the resistivity  gets  the power of $T$ solely from the contribution of DoS, which is linear in $T$. 
			Therefore, in this model, the unusually fast relaxation of strange metal originates from the     correlation of spatially dependent random couplings. Whether such random correlation can be attributed to the strong coupling nature of the material is not clear. 
\end{enumerate}

A summary of three models discussed in this section is given in Table \ref{table:summary0}. Since all these models include impurities, the table only shows the dependence on temperature from various interaction types.
\begin{table}[htbp]
	\centering
	\begin{tabular}{|c|c|c|c|}
		\hline
		\textbf{scattering type}&\textbf{DoS}&$\bm{1-\mathrm{cos}\theta}$&\textbf{resistivity}\\
		\hline
		$u\psi^{\dagger}\psi^{\dagger}\psi\psi$ &$T^2$&$T^0$ & $  T^{2}$\\ \hline
		$g_{ijk}\phi_i\psi_j^{\dagger}\psi_k$&$T$&$T$ & $  T^2$ \\ \hline
		$g_{ijk}(\br)\phi_i\psi_j^{\dagger}\psi_k$&$T$& $T^0$ & $   T$ \\ \hline
	\end{tabular}
	\caption{ Temperature dependence from DoS and $1-\cos\theta$  for each models. We assumed $d=2$ for the Yukawa models,  and  in all cases we   did not mention the    residual  resistivity $\rho_0$ from impurities.} \label{table:summary0}
\end{table}

Based on the discussion above, eqn.\eqref{eqn:doscos} gives  
\begin{eqnarray}\label{eqn:resis}
	\rho\sim\left\{
	\begin{aligned}
		T^{(d+2)/\alpha}, &\quad \text{without the spatial dependence of the random coupling  }\\
		T^{d/\alpha}, &\quad\text{with  the spatial dependence of the random coupling}
	\end{aligned}
	\right.
\end{eqnarray}
for electron-boson scatterings. So far, all the strange metal comes from the case with $d=2=\alpha$ with spatial disorder. 
 One may obtain linear resistivity from a theory without spatial disorder if 
 $\alpha=d+2$.  
 Such a theory is plausible only if one can get the boson self-energy $\Pi \sim \Omega/q^d$. 
 Construction of such a theory is  not a trivial task and is left for future work. 

 \paragraph{ The \^Role of potential disorder}
The previous subsection shows that the spatial randomness in Yukawa type couplings alters a quadratic resistivity to a linear one. Here we briefly discuss the \^role of potential disorder $v_{ij}(\br)\psi^{\dagger}_i\psi_j$.
A more detailed summary of various models is presented in Appendix \ref{App:summary}\\

Typically, the potential disorder brings impurities to a model, resulting in a constant residual resistivity $\rho_0$ \cite{Patel:2022gdh,Guo2022,Hartnoll_Lucas_Sachdev_2018}. Its r\^ole, however, is more than contributing a constant resistivity. 
In \cite{Kim:1994tfo}, the authors find that for a spatially uniform clean model characterized by $g\psi^{\dagger}\psi\phi$ (or $g_{ijl}\psi_i^{\dagger}\psi_j\phi_l$ in \cite{Esterlis2021}), where an FS coupled to a $U(1)$ gauge field without potential disorder, there would be a resistivity $\rho\sim T^{4/3}$ from electron-boson scatterings. If one turns on the potential disorder, the propagator and self-energies will change, so the resistivity from electron-boson scatterings becomes $\rho\sim T$ to the lowest order \cite{Guo2022}. However, it was shown in \cite{Guo2022,DariusShi2022,Shi2024,Gindikin_2024} that such a resistivity ($\rho\sim T^{4/3}$ or $\rho\sim T$) is \emph{zero} as consequence of cancellations among the diagrams. \\

For the spatially random Yukawa model $g_{ijl}(\br)\psi^{\dagger}_i\psi_j\phi_l$, the qualitative behavior is almost the same after one turns off the potential disorder. Without $v_{ij}(\br)\psi_i^{\dagger}\psi_j$, the electron-boson scattering still yields a linear-$T$ resistivity \cite{Esterlis2021}, except that there will be no residual resistivity $\rho_0$.

\section{Conclusion and Outlooks}\label{sec:conclusion}
In this paper, an FS coupled to a vector field is studied, where the couplings are characterized by a spatially random disorder. One achievement of this paper, as we have seen in section \ref{sec:2d} and section \ref{sec:3d}, is that we verify the r\^ole of the spatial random coupling between electrons and bosons is independent of the boson types. In two-dimensional space, it leads to the strange-metal behavior at low temperatures.  When the model is  generalized to $(3+1)$D, the resistivity is no longer linear in temperatures. This indicates that at least in $(2+1)$-dimensional systems, the spatially random fermion-boson couplings can be a master key for linear-$T$ resistivity. \\

Based on the computation, we tried to find an intuitive way to interpret the linearity caused by spatially random couplings. It has two effects, which work together to bring a system a linear resistivity. The delta function over the coordinate space relaxes the momentum conservation of a scattering process, which changes the self-energies and removes the small-angle corrections. Consequently, the scattering rate will be only proportional to the number of bosons.  Since the boson propagator peaks at $\Omega_{\bq}\sim \bq^2$, the  DoS  $\sim T$, i.e. the resistivity is linear in $T$. Moreover, we find two possible ways  leading to linear-$T$ resistivity: one by making $\Omega_{\bq}\sim \bq^{d+2}$ without spatially random coupling and the other by making $\Omega_{\bq}\sim\bq^{d}$ with spatially random coupling. It depends on whether the polarization receives vertex corrections or not. It is of future interest to build various candidates for strange metal according to these criteria.\\

One may ask what the origin of the vector field appearing in this work is. 
In fact, vector fields such as gauge fields coupled with fermions have been widely investigated  \cite{PhysRevLett.65.653,PhysRevB.47.7312,Kim:1994tfo,Polchinski:1993ii,PhysRevB.50.14048,PhysRevB.104.035140,PhysRevLett.122.133602,PhysRevLett.130.083603}. Such a system is related to fractional quantum hall effects and high-temperature superconductivity \cite{Kim:1994tfo}. 
The $U(1)$ gauge field can be either an emergent one or simply an electromagnetic field.  This paper focuses on a vector field instead of a gauge field, and the origin of this field is not important, since the emphasis is to check the universality of the method suggested in \cite{Patel:2022gdh} and to show that it is independent of the interaction type. \\

The most tantalizing question left is whether the realization of random spatial disorder has any connection to the strongly coupled system regarding its origin.  While all the calculations  in this work as well as those in  \cite{Patel2023}   were performed with the assumption of weak coupling, the actual material showing the strange metal are all strongly correlated ones. Without such connections, the mechanism  would be  still remote to the nature. \\

There are still lots of aspects to be studied before a rigorous theory of strange metal is constructed. It is also interesting to investigate the scaling of critical temperature of superconductivity in this type of models. Recently, it has been  observed that at low temperatures, if a system has a resistivity $\rho=\rho_0+AT$, then its critical temperature can satisfy $T_c\propto A^{1/2}$ \cite{Yuan_2022,jiang2023interplay}. As is shown in this paper and in \cite{Patel2023}, $A$ is proportional to $K^2$ (or $g^2$), which is the coupling constant between fermions and bosons. This provides a testing ground for the theory described by eqn.(\ref{eqn:action1}). It has been found in spatially random Yukawa model, $T_c\sim A$ instead of $A^{1/2}$ \cite{Li:2024kxr}.
If one can find that in vector models, $T_c\sim K$, then the reliability of our model can be further enhanced. \\

Another aspect to explore is to flesh out the model (\ref{eqn:action3}) with more physical details. 
For example, 
 the presence of an external magnetic field can be investigated. Here we only study the conductivity of the system, but in addition to the linear-$T$ resistivity, a Hall angle $\sim T^2$ is another important feature of cuprate strange metals \cite{Blake:2014yla,Schrieffer_Brooks_2014}. Therefore, the behavior of Hall angle provides a criterion to verify if the model (\ref{eqn:action3}) describes certain types of strange metals.\\

Finally, holography has turned out to be a powerful tool for studying the condensed matter systems \cite{Herzog:2009xv,Hartnoll:2009sz,McGreevy:2009xe}. 
For instance, the linear-$T$ conductivity as well as the quadratic-$T$ Hall angle can be realized by a black hole \cite{Ge:2016sel,Ge:2016lyn}. In \cite{Huang:2023ihu}, the authors use holographic formalism to study the quantum critical points. Therefore, developing the holographic description and exploring phenomenology can be an interesting topic in the future. \\

\acknowledgments
The authors would like to thank Yunkyu Bang, Yu-Ge Chen, Han-Yong Choi, Yi Zhang, and Difan Zhou for the insightful discussion. 
This work is partly supported by NSFC, China (Grant No. 12275166 and No. 12311540141). This work is also supported by  NRF  of Korea with grant No. NRF-2021R1A2B5B02002603, RS-2023-00218998 and NRF-2022H1D3A3A01077468.

\appendix

\section{A  brief  Summary of  a few Models}\label{App:summary}

This appendix is a complement of section \ref{sec:discussion}. We present a brief review on several relevant models, and generalize Bloch's argument to these models as well.\\

This paper following the line of \cite{Patel:2022gdh} shows that spatially random interactions offer one approach to the theory of strange metal. In section \ref{sec:discussion}, we interpret the linear resistivity by simply counting DoS and small-angle correction via eqn.\eqref{eqn:doscos}.
Now let us delve into the underlying mechanism, illustrating the effects of the random spatial disorder with more detail. To this end, we first summarize the contents of \cite{Esterlis2021,Guo2022,Patel:2022gdh}. Since the FS coupled to the scalar and the FS coupled to the vector share the same qualitative properties, we take models based on the scalar type Yukawa coupling for simplicity. \\

Now we consider the effects of potential disorder $v(\br)\psi^{\dagger}(\br)\psi(\br)$  without flavor indices \cite{Altland2023}. Usually one takes $\langle v(\br)\rangle=0$ and $\langle v^*(\br)v(\br')\rangle=v^2 f(\br-\br')$. Its first-order contribution to the  resistivity is  ${\cal O}(T^0)$\footnote{The introduction of flavor in the disorder potential $v(\br)\to v_{ij}(\br)$ hardly changes the property, so we will not think of $v_{ij}(\br)\psi^{\dagger}_i(\br)\psi_j(\br)$  separately. }.
Generally, there exists vertex corrections \cite{Coleman2019}, and the total resistivity can be evaluated from the following Feynman diagram,
\begin{equation}
	\includegraphics[scale=0.45]{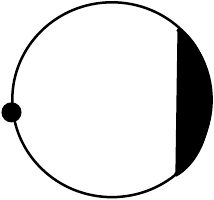}\,,
\end{equation}
which contains an angular average.
If we take $f(\br-\br')\sim\delta(\br-\br')$, there is no vertex correction \cite{Patel:2022gdh}, so the total polarization has a simple form given in \eqref{eqn:v-polar}. 
In addition, a potential $v_{ij}\psi^{\dagger}_{i}(\br)\psi_j(\br)$, which is  disordered only in flavor space without spatial dependence with zero mean, has no contribution, because it leaves the Fermion propagator unchanged.\\

Let us continue to the interaction between the fermions and a vector field in $(2+1)$ dimension. 
The computation in \cite{Kim:1994tfo}, where 
the Fermi surface coupled with a $U(1)$ gauge field without disorder was studied, provides us a guide  to study this type of models \cite{Guo2022}.
According to ref.\cite{Kim:1994tfo}, a part of MT diagram \eqref{graph:mt} cancels the self-energy contribution, and the rest part of MT diagram, together with AL diagrams \eqref{graph:al1} should yield a resistivity $\rho\sim T^{4/3}$. \\

The scaling behavior is further verified in \cite{Guo2022}, where a random Yukawa term $g_{ijk}\phi_i\psi^{\dagger}_j\psi_k$ is studied \emph{without} potential disorder, so it also named a clean random Yukawa model. In this theory, the random disorder is introduced. However, the color disorder does not change the fundamental properties of the model, and the authors also find the contribution from MT and AL diagrams would result in a resistivity $\rho\sim T^{4/3}$.\\

So it seems that the total contributions result in a conductivity $\sigma^{xx}\sim T^{-2/3}$, or a resistivity $\rho\sim T^{4/3}$ \cite{Kim:1994tfo}. However, the calculation in \cite{Kim:1994tfo} is corrected by \cite{Guo2022} such that the summation of all diagrams is zero. In other words, the coefficient of $T^{4/3}$ is $0$. 
The absence of resistivity from electron-boson scattering is a result of ``boson drag'' \cite{Patel:2022gdh,Chowdhury2022,Guo2022}, as will be illustrated later. One main conclusion here is that the random coupling alone cannot lead to a theory of strange metal.
\\

As mentioned in section \ref{sec:discussion}, after introducing a spatial disorder potential $v_{ij}(\br)\psi^{\dagger}\psi$,  it was shown  in \cite{Guo2022, Patel:2022gdh} that the electron self-energy reads $\omega\ln(1/\omega)$. The presence of impurity widens the peak at the Fermi surface \cite{Guo2022}, and this brings a overwhelmingly large scattering rate $\Gamma$ which is of order $\omega^0$. Therefore, in the computation of Dyson's equations, we can drop many higher-order terms in $\omega$. The computation shows that  the  Feynman graph
\begin{equation}\label{graph:conduct}
	\includegraphics[scale=0.45]{eself-energy}\,
\end{equation}
contributes to the conductivity a term linear in temperature. However, such linear-$T$ dependence is precisely canceled by the MT diagrams (\ref{graph:mt}) \cite{Patel:2022gdh,Guo2022}. 
Due to this cancellation, the higher order AL diagrams finally brings a resistivity $\rho\sim T^2$ \cite{Guo2022}. As a consequence, there is no strange-metal behavior either in this theory \cite{Patel:2022gdh,Guo2022}. Despite the failure,  the theory  achieved   the realization that the spatially random disorder is very likely the steppingstone to the strange metal. \\

Finally, the authors of \cite{Patel:2022gdh} find a way to get rid of the cancellation of MT graphs. The key is the spatial dependence of the coupling $g_{ijk}(\br)$  in the   interaction  term $g_{ijk}(\br)\phi_i\psi^{\dagger}_j\psi_k$, which satisfies $$\langle g^*_{ijk}(\br)g_{i'j'k'}(\br')\rangle=g^2\delta(\br-\br')\delta_{ii'jj'kk'}.$$
Because of the spatial delta, the MT diagram and AL diagrams vanish, and the only contribution is from eqn.(\ref{graph:conduct}) so that the terms  linear-$T$ survives. Additionally, in \cite{Esterlis2021}, the authors find that the spatially random Yukawa coupling can yield a linear-$T$ resistivity even without the potential disorder.\\

In short, the flavor disordered Yukawa(-like) couplings hardly change the physical properties, but the \emph{spatial disorder} is the key to obtaining the strange-metal behavior, as is summarized in table \ref{table:summary}, where we have compared various typical electron-boson scattering models.\\

\begin{table}[htbp]
	\centering
	\begin{tabular}{|c|c|}
		\hline
		\textbf{scattering type}&\textbf{resistivity}\\
		\hline
		$v_{ij}(\br)\psi^{\dagger}_i\psi_j$ & $\sim \rho_0$\\ \hline
		$v_{ij}\psi^{\dagger}_i\psi_j$ & no contribution\\ \hline
		$u\psi^{\dagger}\psi^{\dagger}\psi\psi+$impurities & $\sim \rho_0+AT^{2}$\\ \hline
		$g\phi\psi^{\dagger}\psi$ & $\cancel{T^{4/3}}$ absence\\ \hline
		$g_{ijk}\phi_i\psi_j^{\dagger}\psi_k$ & $\cancel{T^{4/3}}$ absence\\ \hline
		$g_{ijk}\phi_i\psi_j^{\dagger}\psi_k+v_{ij}(\br)\psi^{\dagger}_i\psi_j$ & $\sim \rho_0+\cancel{AT} +B T^2$ \\ \hline
		$g_{ijk}(\br)\phi_i\psi_j^{\dagger}\psi_k$ & $\sim T$\\ \hline
		$g_{ijk}(\br)\phi_i\psi_j^{\dagger}\psi_k+v_{ij}(\br)\psi^{\dagger}_i\psi_j$ & $\sim \rho_0+AT$ \\ \hline
	\end{tabular}
	\caption{A comparison among models with and without spatial random couplings. } \label{table:summary}
\end{table}

Now let us continue to the mechanism for the linear-$T$ resistivity.
One may wonder why the linear-$T$ resistivity is so hard to get and why it is obtained only in spatial dimensions $2$. This subsection will offer some intuitive discussion following section \ref{sec:discussion}. The electron-boson scattering described by Fig.\ref{fig:ii} yields a resistivity. The resistivity consists of contributions from DoS of bosons, $T^{d/\alpha}$, and small-angle contributions, $T^{2/\alpha}$.\\

Now let us test the models mentioned above. The relation $\epsilon_{\bq}$ can be derived from the boson propagator $\Pi(\bi\Omega,\bq)$. In both regular Yukawa  model \cite{Kim:1994tfo} and the clean random Yukawa \cite{Esterlis2021}, 
the boson self-energy is given by \cite{Guo2022}
\begin{eqnarray}
	\Pi(\bi \Omega_m,\bq)
	=-c_g\frac{|\Omega_m|}{|\bq|}, 
\end{eqnarray}
where $c_g$  is a constant. Thus the boson propagator is
\begin{eqnarray}\label{eqn:b-propagatora}
	D(\bi\Omega_m,\bq)\simeq\frac{1}{\bq^2+c_g\frac{|\Omega_m|}{|\bq|}}.
\end{eqnarray}
This implies that the pole of the boson propagator comes from 
\begin{equation}
 \Omega_{ { m}} \sim \bq^3,
\end{equation}
i.e. $\alpha=3$. According to \eqref{eqn:dos}, it gives DoS contribution to the scattering rate  $T^{d/3}$.
 Similarly,  the  small-angle scattering gives $T^{2/3}$.
As a result, the total transport scattering rate depends on $T^{(d+2)/3}$. Taking $d=2$, one recovers the resistivity in a Yukawa-like theory, where $\rho\sim T^{4/3}$ \cite{Kim:1994tfo}. However, as is shown in \cite{Guo2022,DariusShi2022,Shi2024,Gindikin_2024}, this would-be  $T^{4/3}$-dependence turns out to  contribution from other diagrams.\\

In systems without impurities, there exists \emph{`boson drag'}, resulting in an absent resistivity \cite{Patel:2022gdh,Guo2022,Chowdhury2022}. Though following Bloch's argument, one should find the resistivity satisfies $\rho\sim T^{4/3}$, Peierls argues that ``phonon drag'' will make the resistivity decrease faster than one expects from eqn.\eqref{eqn:doscos} \cite{Peierls1930,peierls1932,Ashcroft_Mermin_1976}. When temperatures are low enough such that the umklapp process cannot happen, the \emph{total} momentum of bosons and fermions is conserved in a clean model. The decay of the electric current becomes infinitely slow, so, as is shown in \cite{Guo2022}, the scattering is unable to contribute and the resistivity vanishes. Therefore, in models with global momentum conservation, eqn.\eqref{eqn:doscos} is not appropriate for achieving resistivity, and we should expect a vanishing resistivity due to the ``boson drag'' \cite{Chowdhury2022}.
Conservation of the total momentum is necessary for a boson drag to happen, so for the time being, phonon drag is only observed in very pure sample \cite{Patel:2022gdh,PhysRevLett.109.116401,Chowdhury2022}. In the presence of impurities, the total momentum will decline and there will be no boson drag.\\

Now let us see how the result is changed by introducing the potential disorder $v(\br)\psi^{\dagger}\psi$ \cite{Guo2022}.  
Such an impurity potential not only prevents the system from boson drag by breaking the global momentum conservation, but also alters the behavior of the boson propagator. 
Because the self energy $\Pi$ becomes $\Omega/\sqrt{\Gamma^2+q^2}$ 
and $\Gamma \gg q$,  
the boson propagator reads
\begin{equation}\label{eqn:b-propagatorbb}
	D(\bi\Omega_m,\bq)\simeq\frac{1}{\bq^2+c_0|\Omega_m|}.
\end{equation}
%
%
Then the poles are along $\epsilon \sim \bq^2$ so that dispersion relation changes drastically. 
%
With $\alpha=2$, one finds $\rho\sim T^{(d+2)/2}$ according to eqn.\eqref{eqn:doscos}. Choosing $d=2$, one finds the result in \cite{Guo2022} with a resistivity $\sim T^2$ due to the vertex corrections from AL diagrams.\\

As summarized in section \ref{sec:discussion}, there is no correction from $(1-\cos\theta)$ in spatially random Yukawa model. Only DoS begets a $T^{d/2}$ resistivity. Moreover, the spatial randomness dropped the momentum conservation on vertices, and the global momentum conservation is broken as well. Therefore, even without impurities, a boson drag is impossible to happen in a clean spatially random system \cite{Esterlis2021}. Hence, when $d=2$, the spatially random Yukawa coupling yields a resistivity linear in $T$. 
While in $(3+1)$ dimensions, this property disappears, and one finds $\rho\sim T^{3/2}$. This result of $(3+1)$ dimensions is not only consistent with \cite{Patel:2022gdh}, but also with section \ref{sec:3d} in this article.
Furthermore, based on this analysis, one finds that the result is qualitatively the same regardless of the type of bosons (scalar or vector) that the electrons are coupled to, since only the relation $\epsilon(\bq)$ and spatial dimensions contribute.\\

In essence, eqn.\eqref{eqn:resis} can be further modified into table \ref{table:resistivity}. Suppose we have a system in $(d+1)$-dimensional spacetime, where an FS is coupled to critical bosons. In order that the scattering has non-trivial contributions to the resistivity, the first step is to break the total momentum conservation of bosons and electrons. This can be achieved by either introducing impurities (such as potential disorder) or spatial disorder. Then the resistivity $\rho\sim T^{(d+2)/\alpha}$ for spatially uniform couplings, and $\rho\sim T^{d/\alpha}$ for spatially random couplings.
\begin{table}[htbp]
	\centering
	\begin{tabular}{|c|c|c|}
		\hline
		\thead{\textbf{spatial}\\\textbf{disorder}}&\thead{\textbf{global momentum}\\\textbf{conservation}}&\textbf{resistivity}\\
		\hline
		$\times$ &$\surd$& absent\\ \hline
		$\times$ & $\times$ &$T^{(d+2)/\alpha}$ \\ \hline
	    $\surd $ & $\times $ & $T^{d/\alpha}$\\ \hline
	\end{tabular}
	\caption{The resistivity from electron-boson scatterings.} \label{table:resistivity}
\end{table}

\bibliographystyle{JHEP}
\bibliography{disorderedQED.bib}


\end{document}